%% file: 0-main.tex
\documentclass[conference]{IEEEtran}

\usepackage{cite}
\usepackage{amsmath,amssymb,amsfonts}
\usepackage{graphicx}
\usepackage{textcomp}
\usepackage{xcolor}

\usepackage{url}
\usepackage[hidelinks]{hyperref}
\usepackage[normalem]{ulem}
\usepackage{booktabs}
\usepackage{multirow}
\usepackage{tabularx}
\usepackage{threeparttable}
\usepackage{tablefootnote}
\usepackage{float}
\usepackage{caption}
\usepackage{subcaption}
\usepackage{svg}
\usepackage{colortbl}
\usepackage[ruled,linesnumbered]{algorithm2e}
\usepackage[capitalize,noabbrev]{cleveref}
\usepackage{tikz}
\usepackage{pgfplots}
\pgfplotsset{width=10cm,compat=1.9}
\usepackage{listings}
\usepackage{xspace}
\usepackage[many]{tcolorbox}
\usepackage[compact]{titlesec}

\makeatletter
\g@addto@macro\normalsize{%
  \setlength{\abovedisplayskip}{4pt}%
  \setlength{\belowdisplayskip}{4pt}%
  \setlength{\abovedisplayshortskip}{2pt}%
  \setlength{\belowdisplayshortskip}{2pt}%
}
\makeatother

\titlespacing*{\section}
  {0pt}{0.8ex plus 0.2ex minus 0.2ex}{0.6ex plus 0.2ex}

\titlespacing*{\subsection}
  {0pt}{0.6ex plus 0.2ex minus 0.2ex}{0.4ex plus 0.2ex}

\titlespacing*{\subsubsection}
  {0pt}{0.5ex plus 0.2ex minus 0.2ex}{0.3ex plus 0.2ex}

\definecolor{LightGray}{gray}{0.92}
\definecolor{RowGray}{gray}{0.95}
\definecolor{GrayText}{gray}{0.3}
\definecolor{lightgray}{gray}{0.95}
\definecolor{rowBlue}{rgb}{0.8,0.9,1}
\definecolor{darkgreen}{rgb}{0.0, 0.5, 0.0}

\newcommand{\name}{\textsc{ReBug}\xspace}

\begin{document}

\title{
% From Bug Reports to Failures: An LLM-Driven Agent for Automated Reproduction of Web GUI Bugs
From Bug Reports to Browser-Executable Procedures: An LLM-Driven Agent for Web GUI Bug Reproduction
}

\author{
\IEEEauthorblockN{Cunming Zhang}
\IEEEauthorblockA{
University of Luxembourg\\
cunming.zhang@uni.lu
}

\and

\IEEEauthorblockN{Yu Pei}
\IEEEauthorblockA{
University of Luxembourg\\
yu.pei@uni.lu
}

\and

\IEEEauthorblockN{Michail Papadakis}
\IEEEauthorblockA{
University of Luxembourg\\
michail.papadakis@uni.lu
}
}

\maketitle

\begin{abstract}
% We present \name, a context-aware agent system that reproduces web GUI bugs directly from bug reports by driving a real browser.
% Our results show that combining explicit context reconstruction with state-aware browser execution can automate approximately half of web GUI bug reproduction in realistic settings.

Reproducing web GUI bugs from natural-language bug reports is critical for software maintenance, but remains difficult because reports often lack prerequisites such as dependencies and input files. Existing bug reproduction techniques mainly target code units or mobile applications and lack end-to-end visual execution and validation for web GUIs. We present \name, a context-aware agent system that reconstructs, executes, and validates browser-level reproduction procedures from web GUI bug reports by driving a real browser. \name separates reproduction into two stages. In the preparation stage, \name reconstructs missing prerequisites from the report and available artifacts, and it produces a high-level reproduction plan. In the execution stage, it performs tool-mediated interactions in the browser, maintains structured summaries of page state and action history, and validates the final state against expectations derived from the report.
We evaluate \name on 667 real-world bug reports from four open-source web applications. On controlled current deployments, \name outperforms both baselines, achieving an average RSR of 49.96\%, a mean task completion rate of 74.96\%, and a mean action execution success rate of 86.54\%. Our results show that explicit context reconstruction and state-aware browser execution effectively support report-derived browser reproduction, while historical replay shows that successful procedures often expose the original bug-present behavior on restored buggy versions.
\end{abstract}


\input{1-Introduction}
\input{2-Motivation}
\input{3-Approach}

\input{4-Study_design}
\input{5-Results}
\input{5.1-discussion}
\input{6-Threats}
\input{7-Related_work}
\input{8-Conclusion}

% \section*{Data Availability}
% To preserve double-blind review, the dataset, code, and replication package are anonymized and will be released upon acceptance.

\bibliographystyle{IEEEtran}
\bibliography{9-References}

\end{document}

%% file: 1-Introduction.tex
 \section{Introduction}\label{sec:introduciton}

Reproducing web GUI failures from bug reports is a critical prerequisite for software testing and maintenance, as it allows developers to observe reported failures before diagnosing their causes~\cite{roehm2013monitoring, burg2013interactive}. In practice, reproduction requires translating an informal description into a concrete sequence of browser interactions and checking whether the observed outcome is consistent with the reported behavior. Despite its importance, reproduction remains largely manual and labor intensive, as developers must interpret informal natural language descriptions and translate them into executable UI actions.

Modern web applications make this task difficult for several reasons. First, bug reports often omit important reproduction information, such as entry points, authentication requirements, environmental conditions, or test data~\cite{bettenburg2008makes, chaparro2019assessing}. Second, web GUIs, especially single-page apps are highly dynamic: page states depend on user context, navigation history, asynchronous events, and changing DOM structures~\cite{herbold2011improved}. Third, the growing volume of bug reports makes manual reproduction costly at scale~\cite{AnvikHM06}. These challenges motivate automated support for transforming natural-language reports into executable web GUI reproduction procedures.

Existing techniques provide only partial support for this setting. Browser automation frameworks such as Selenium and Playwright offer the execution base for web interactions~\cite{gojare2015analysis, pathak2024web}, but they assume that selectors and test logic are already available and can be brittle under locator or DOM changes~\cite{brisset2022erratum}. Prior bug reproduction research has made progress in mobile and code-level settings~\cite{kang2023large, khatib2025assertflip}. Android systems such as RepRev, ReCDroid+, AdbGPT, and ReBL extract information from reviews or bug reports, match it to Android GUI elements, or use feedback during replay~\cite{li2020automated, zhao2022recdroid+, feng2024prompting, wang2024feedback}. Existing implementations and evaluation assumptions in these systems are built around mobile-specific substrates such as APKs, Android UI hierarchies, platform instrumentation, and mobile-oriented failure signals. Web GUI reproduction must instantiate these ideas through a general browser, where actions and outcomes are grounded in browser-observable evidence such as URLs, DOM snapshots, screenshots, tab states, and asynchronous page updates. Existing web-focused work has often addressed GUI testing, exploration, or runtime recording, while automated reproduction from natural-language bug reports remains less explored~\cite{wang2024application, chen2025standing}.
% However, these systems rely on Android-specific execution structures, UI hierarchies, instrumentation, or mobile-oriented failure signals, which do not directly transfer to browser-based web GUIs. We do not claim that the underlying algorithmic ideas of these systems, such as guided exploration, state tracking, or feedback-driven replay, are inherently limited to Android. Rather, their available implementations and evaluation assumptions are built around mobile-specific execution substrates and signals, whereas web GUI reproduction must instantiate these ideas over browser-level observations, dynamically changing DOMs, URL and tab states, screenshots, web artifacts, and asynchronous client-side updates.
% Unlike native mobile reproduction, web GUI reproduction must operate through a general browser and ground actions in browser-observable states such as URLs, DOM snapshots, screenshots, and asynchronous page updates. 

% Recent LLM-based web agents and benchmarks show that language models can interpret instructions, act in realistic browser environments, incorporate visual grounding, and interleave reasoning with actions~\cite{deng2023mind2web, zhou2023webarena, zheng2024gpt, yao2022react}. However, they mainly target task completion or test generation, whereas bug reproduction additionally requires completing missing execution context, tracking state during interaction, and validating the final browser-observable outcome against the report-derived oracle.

Recent LLM-based web agents and benchmarks show that language models can interpret instructions and interact with browser environments~\cite{deng2023mind2web, zhou2023webarena, zheng2024gpt, yao2022react}. Nevertheless, these agents have a primary focus on task completion or test generation, whereas bug reproduction necessitates supplementary capabilities, such as the ability to recover missing execution context, adapt actions to dynamic browser states, and validate the final browser-observable outcome against report-derived expectations. Taken together, existing approaches only partially address web GUI bug reproduction, as they often assume fixed scripts, generic task execution, or Android/code-level execution models that do not fully capture missing context, dynamic browser states, and browser-observable outcomes.

% To address this gap, we leverage the natural language bug reports not only as documentation artifacts but also as actionable inputs for automated testing. We present \name, a novel context aware, LLM driven agent designed to perform end-to-end reproduction of web GUI bugs directly from natural language bug reports. \name follows a staged pipeline that infers missing reproduction context from reports and project artifacts, executes parameterized browser interactions, and evaluates outcomes by comparing observed runtime states against report-derived expectations. Given a bug report, \name navigates the target web application, executes inferred user interactions, and observes runtime behaviors to reconstruct the reported failure scenario. Through this process, \name aims to determine whether the report-described reproduction procedure can be reconstructed. This design allows \name to infer missed prerequisites, reason over intermediate UI states during execution, and validate reproduction outcomes against report derived specifications. These capabilities are not jointly supported by existing LLM based web agents or testing frameworks.

To address this gap, we present \name, a context-aware LLM agent that reconstructs and validates browser-level reproduction procedures from natural-language web GUI bug reports. \name first prepares report-specific execution context when needed, then drives a real browser with state-aware actions, and finally checks the resulting browser-observable state against report-derived expectations.

We evaluate \name on 667 real world web GUI bug reports collected from the GitHub issue trackers of four open source web applications. To account for report quality and software evolution, we curate the benchmark through a multistage construction pipeline: 
(i) large scale collection and confirmation reduce thousands of bug issues to a set of confirmed bug reports; (ii) refinement removes low quality or environment incompatible reports; and (iii) annotation specifies outcome oracles in terms of browser-observable signals, enabling consistent automated evaluation of reproduction outcomes.

Across the benchmark, \name achieves an average reproduction success rate of 49.96\% and demonstrates consistent improvements over baselines in task completion and action level execution. Since the large-scale evaluation mainly measures report-derived procedure reconstruction and current-deployment outcome validation, we further study 40 successful cases through historical replay and analyze task completion, action execution, cost, and failure modes to identify the strengths and limitations of automated web GUI reproduction. The results suggest that explicit context reconstruction and state-aware browser execution are useful in this setting, while dynamic UI grounding and missing prerequisites remain important challenges.

%In addition to effectiveness, we analyze cost and failure modes by measuring LLM calls, wall clock time, and the distribution of failure categories derived from execution traces. 

% Overall, these findings indicate that \name can support developers, testers, and automated debugging pipelines by enabling more reliable reproduction of web GUI failures across diverse web applications.

In summary, this paper makes the following contributions:
\begin{itemize}
  \item We propose \name, a context-aware, LLM-driven agent that integrates prerequisite reconstruction, state-aware browser execution, and report-derived outcome checking to construct and execute browser-executable reproduction procedures from natural-language web GUI bug reports.
  
  \item We conduct an empirical evaluation on 667 real-world web GUI bug reports and compare \name with direct script generation and a browser-agent baseline. We further perform a historical replay study on 40 successful cases, providing complementary evidence that reconstructed procedures can trigger the report-described failures when the corresponding buggy versions are restored.
  
  \item We build a benchmark of real-world GitHub web GUI bug reports with annotated outcome oracles, supporting systematic evaluation and future research on LLM-based testing and web GUI bug reproduction.
\end{itemize}

%% file: 2-Motivation.tex
\section{Motivation}\label{sec:motivation}

Reproducing web GUI bugs from bug reports (written in natural language) is necessary for understanding failures, validating fixes, and supporting regression testing in modern web applications. Empirical studies show that developers often encounter difficulties reproducing reported bugs due to missing execution steps, incomplete environment information, and the nondeterminism of complex systems~\cite{lota5060080recent}. These challenges highlight the need for more systematic support for bug reproduction.
Existing web testing frameworks and browser agents provide useful automation support, but they typically assume predefined scripts, stable interaction goals, or manually specified preconditions. In contrast, bug reproduction starts from an informal report and requires constructing an executable procedure whose final browser state can be checked against the reported behavior.

From a systems perspective, three technical limitations remain insufficiently addressed. 
\begin{itemize}
  \item \textbf{Context Completion:} Reports may depend on prerequisites that are only partially specified, such as configuration, data, or files. Existing approaches provide limited support for preparing such execution context from the report and project-level artifacts.

  \item \textbf{State-aware Execution:} Web interfaces may change during interaction due to navigation, asynchronous updates, or dynamic DOM changes. The same action may lead to different results depending on the current page state and previous steps, requiring execution decisions to be grounded in recently observed browser state.
  
  % and the same action can have different outcomes depending on the current page state and prior steps. Automated interaction tools often execute actions without maintaining a compact and structured representation of what has been observed so far, which makes it harder to reason about intermediate states and recover from UI changes during long procedures.

  \item \textbf{Outcome Validation:} In many settings, success is decided implicitly or through manual inspection. Without a consistent comparison between observed browser states and report implied expectations, it is difficult to judge whether a run truly reproduces the reported outcome.
\end{itemize}

Together, these limitations motivate a staged treatment of web GUI bug reproduction. Although the underlying challenges are common in bug reproduction, the web setting makes them operationally different: the system cannot rely on APK-level instrumentation, mobile UI hierarchies, or platform crash signals, and must instead reason from browser-observable DOM states, screenshots, URLs, asynchronous updates, and GUI-consumable artifacts.

% Together, these limitations motivate treating web GUI bug reproduction as a staged process rather than a single script-generation or task-completion problem. These challenges are not unique to web GUIs; they are general difficulties in automated bug reproduction. The web setting makes them operationally different because the system cannot rely on mobile UI hierarchies, APK-level instrumentation, or platform crash signals, and must instead make reproduction decisions from browser-observable evidence such as DOM states, screenshots, URLs, asynchronous page updates, and GUI-consumable artifacts.

%% file: 3-Approach.tex
\section{Approach}\label{sec:methodology} 

\subsection{Overview of \name}\label{sec:methodology-overview}
\begin{figure*}[t]
  \centering
  \includegraphics[width=\textwidth]{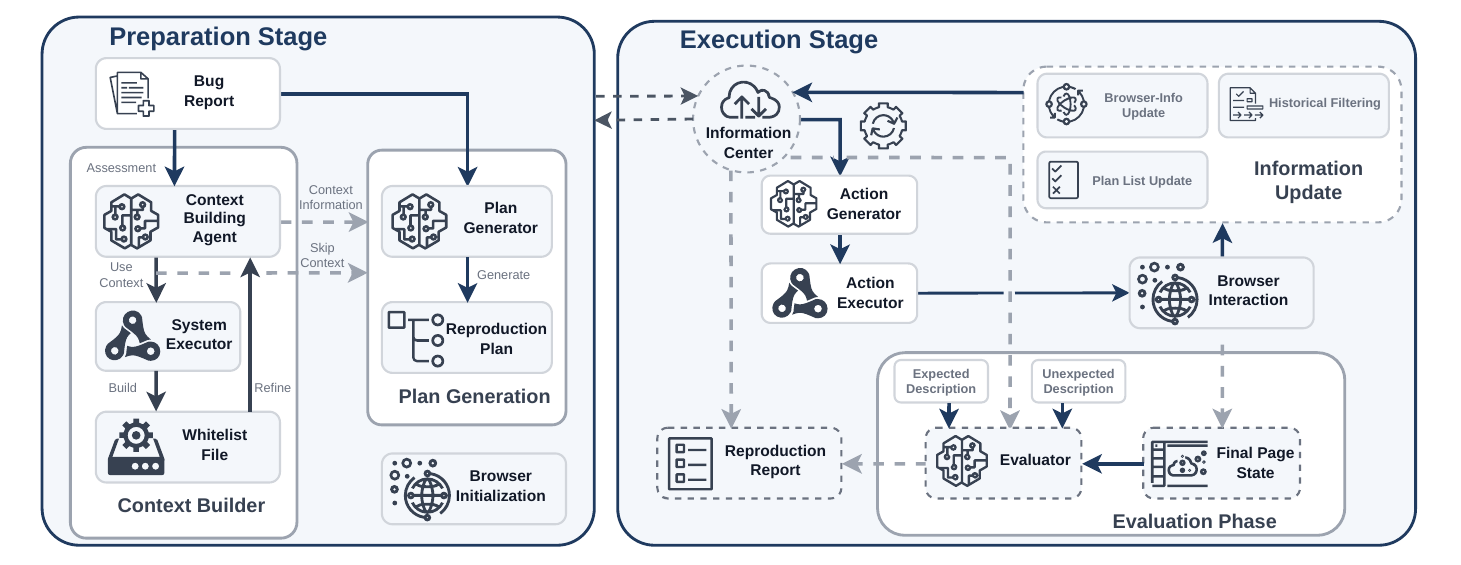}
  \caption{Overview of \name. The workflow begins with the preparation stage, which reconstructs the required context and produces a reproduction plan. It then enters the execution stage, which performs an iterative execution loop in a real browser with continuous state tracking and trace logging, followed by automatic outcome evaluation.}
  \label{fig:overview}
  \vspace{-0.8em}
\end{figure*}

\name constructs, executes, and checks browser-executable reproduction procedures from natural-language web GUI bug reports. Given a report, \name first determines whether the run requires additional resources beyond the baseline setup available in the evaluation environment. These resources include prerequisite data, input files, or other local artifacts needed during execution.
The preparation stage produces an artifact inventory and a high-level reproduction plan. The execution stage uses them to drive a real browser, record a structured trace, and collect the final observed state for oracle checking. Figure~\ref{fig:overview} gives an overview of the workflow.
% \name is a pipeline for constructing, executing, and validating browser-level reproduction procedures from natural-language web GUI bug reports. \name first inspects the report and determines whether additional execution context is required beyond the account information and web entry points manually provided in the evaluation setup, and whether the required context can be prepared in the evaluation environment. Such context mainly refers to artifact-level resources, such as prerequisite data, required files, or other local resources needed for browser-based reproduction. The preparation stage outputs an artifact inventory and a high level reproduction plan, and the execution stage consumes them to perform browser actions, maintain a structured execution trace, and produce a final browser-observable state for validation. Figure~\ref{fig:overview} gives an overview of the workflow.

% \begin{figure}[!h]
% \centering
% \includegraphics[width=0.98\linewidth]{figure/3.1-workflow_v3.pdf}
% \caption{\textbf{Overview of \name.} The workflow begins with the preparation stage, which reconstructs the required context and produces a reproduction plan. It then enters the execution stage, which performs an iterative execution loop in a real browser with continuous state tracking and trace logging, followed by automatic outcome evaluation.}
% \label{fig: workflow} 
% \end{figure}

\subsection{Preparation Stage}
% The preparation stage converts a natural language bug report into an executable setup for reproduction. It identifies whether artifact-level context is needed, prepares such artifacts when feasible, generates a plan based on the report and the prepared artifacts, and initializes the browser from the reported or default entry point. 
The preparation stage operationalizes the setup decisions identified in the overview: it assesses report-specific prerequisites, materializes feasible artifacts, generates a high-level plan, and initializes the browser entry state.
% When the report provides no entry information, it initializes the browser from the application homepage.

\subsubsection{Context Builder}

\textit{\textbf{Phase 1: Assessment.}}
% For each bug report, the Context Builder first performs an assessment to decide whether the report depends on execution prerequisites that must be materialized before browser interaction. In this paper, execution prerequisites refer to report-specific resources or states that are not automatically present in a fresh deployment but are necessary to follow the reported reproduction steps. Examples include files to be uploaded, images or documents mentioned in the report, configuration snippets, spreadsheet-like data, or other local artifacts that the browser procedure needs to consume. They do not include general deployment conditions such as installing the application, switching application versions, rebuilding the database backend, or reproducing the reporter's entire machine environment. The \textbf{Context Building Agent} reads the report and checks whether the described steps rely on such materializable resources. The assessment produces a prerequisite summary that records the required artifact type, its source when available, whether it can be generated or reused locally, and the reason when it cannot be prepared. If the report is self-contained, the system skips artifact preparation and proceeds directly to browser execution. When prerequisites are required, the agent further assesses whether they can be prepared and validated within the controlled execution environment. If the prerequisites cannot be prepared or reliably validated, the system avoids unnecessary construction and records the reason for later analysis.

For each report, the Context Builder determines whether browser execution requires materializable report-specific prerequisites that are absent from a fresh deployment. These prerequisites include files, images, documents, configuration snippets, spreadsheet-like data, or other local artifacts consumed by the browser procedure; they exclude deployment-level conditions such as installing the application, switching versions, rebuilding databases, or reproducing the reporter’s machine environment. The assessment outputs a prerequisite summary recording the artifact type, source, preparation feasibility, local path if prepared, and failure reason if unavailable. Self-contained reports skip artifact preparation, while infeasible prerequisites are recorded for later analysis.

The whitelist defines the local resource boundary available to the Context Builder. It contains only evaluation-provided directories, downloaded report attachments, and artifacts produced by preparation steps for the same report. The agent can also reuse existing local resources by scanning candidate files and deciding whether they satisfy the prerequisites implied by the report, which reduces redundant preparation work. 

\textit{\textbf{Phase 2: Iterative Materialization.}}
If the context assessment indicates that missing prerequisites must be prepared, the Context Builder enters an execution stage that incrementally materializes the required context. This phase is implemented as an iterative \textbf{Context Building Agent} that follows a build, check, and refine loop. At each iteration, the agent reasons over three signals: the bug report, the prerequisite summary produced by assessment, and the current set of prepared artifacts. It then proposes a small set of concrete preparation operations that move the workspace toward an executable reproduction setup.

% The Context Builder realizes these decisions through a tool supported \textbf{System Executor} rather than free form scripting. The executor runs all operations inside a sandboxed workspace and exposes a bounded set of operations that covers typical context needs. In our implementation, this operation set is limited to local artifact operations, including creating directories, writing or editing files, selecting reusable files from the whitelist, downloading report-linked resources when available, validating file existence and readability, and staging files for later browser upload. These operations include creating and organizing directories, generating and editing structured artifacts, downloading resources, performing lightweight validation, and preparing files for upload when the report implies file based interactions. Each executed operation produces explicit outputs, such as newly created files, updated configurations, or validated resources, and the system persistently records them as reusable artifacts.

The Context Builder realizes these decisions through a tool supported \textbf{System Executor} rather than free-form scripting. The executor runs all operations inside a sandboxed workspace and exposes a bounded set of local artifact operations, including creating or editing files and directories, reusing whitelisted files, downloading report-linked resources, validating file existence and readability, and staging files for browser upload. Each executed operation produces explicit outputs, such as newly created files, updated configurations, or validated resources, which are persistently recorded as reusable artifacts.

To keep the process robust and traceable, the system maintains an explicit artifact inventory and an action history. The inventory enables the agent to reuse previously prepared resources and to avoid duplicating work. The action history grounds future decisions and provides evidence for debugging and evaluation, including which preparation step failed, which file was produced, and what validation result was observed. The preparation loop terminates when the Context Builder emits an explicit \texttt{done} action after the requested artifacts have been prepared, or when the preparation budget is exhausted.

\subsubsection{Plan Generation}
% Before browser execution, \name derives a reproduction plan from the bug report and available context information. The plan is an ordered sequence of intent steps that provides a stable execution scaffold, rather than a fixed script. During execution, the agent grounds each planned step to the current page state, dynamically selects target elements, refines subactions, and inserts or skips steps when needed to handle dynamic content and unexpected UI variations. The initial plan is treated as a fixed high-level reference rather than an executable script. During execution, the web agent performs concrete step-level planning based on this plan and the current browser state, dynamically generating and adjusting actions as needed. The original plan itself is not modified; instead, it serves as a progress reference to indicate how far the actual execution has advanced relative to the overall reproduction procedure.

Before browser execution, \name derives an ordered sequence of high-level intent steps from the bug report and prepared context. The plan serves as a progress scaffold rather than a fixed script: during execution, the agent grounds each step in the current browser state, dynamically chooses target elements, refines subactions, and may insert or skip actions to handle UI variation. The original plan is not rewritten; it is used to track how far the live execution has advanced.

\subsubsection{Browser Initialization}
At the beginning of a run, \name initializes a fresh browser session with standardized settings and prepares the executor state for step-wise interaction. \name encodes runtime signals as structured fields in the browser-state record, including the current URL, page title, active tab, tab count, navigation status, dialog or popup events, browser errors, action timeout, element-not-found errors, blocked interactions, downloads, and browser responsiveness, which are provided to the Action Generator after each action to help distinguish successful execution, partial progress, and failures. It also verifies that the session is responsive and that browser focus is properly established. If the bug report specifies an entry point, such as a starting URL, the executor navigates to that page to set the initial state. Otherwise, when the report contains no starting page information, \name defaults to launching from the web application's homepage as the entry state, and then proceeds with plan-guided exploration.

\subsection{Execution Stage}
The execution stage drives the browser according to the prepared artifacts and high-level plan, while continuously observing the live page state and recording a trace for validation and analysis.
% it runs an iterative decision loop that repeatedly observes the current browser state, chooses the next execution step that advances the plan, materializes the step into one or more low-level browser actions through tools, and records a trace of what was observed and performed. This design keeps long interaction procedures grounded in the live page state rather than relying on a fixed script.

% \name reproduces bugs through a step-wise execution loop that couples agent-based decision-making with tool-mediated actions in a real browser. Guided by the reproduction plan produced in the preparation stage, the executor runs a step-wise execution loop with an explicit step budget, action-level timeouts, robustness guards, and trace logging, and then performs automatic outcome validation based on the final observed browser state.

\subsubsection{Execution Loop (\textit{Step-wise execution loop})}
% \paragraph{Runtime initialization and pre-context integration.}
% At the beginning of a run, the executor initializes a browser session and enters a pre-context phase that integrates any previously prepared local artifacts into the run state. When context construction produces usable files, their paths are registered as available resources so that subsequent steps can read from or upload them as needed during execution.
% \paragraph{Automatic entry and high-level reproduction plan.}
% Before entering the main execution loop, the executor optionally extracts an initial URL from the bug report and performs a navigation step to establish the starting state. It then generates a high-level reproduction plan from the report text, producing an ordered list of steps that serves as the task backbone throughout execution. When pre-context artifacts are available, they are summarized and injected into the planning prompt so that the plan explicitly accounts for how these resources should be used.
\name executes the reproduction plan through a step-wise interaction loop with the live web page. At each iteration, the information update module captures a screenshot and a structured DOM representation of the current browser state. \name uses the screenshot as runtime visual evidence, together with the visible DOM, interactable elements, URL, tab state, and latest action outcome, to ground subsequent actions and to support final-state validation. This visual evidence is collected uniformly for every report during browser execution; screenshots or other visual attachments in the original report are treated only as optional input artifacts, not as a prerequisite for visual processing. The Browser-Info Update then converts these raw observations into a compact browser-observable record and writes it to the Information Center as the current state. Historical Filtering keeps the decision context compact by exposing only the latest browser state as grounding evidence, while summarizing recent agent history and saving the full action and browser-state trace for offline analysis. This prevents obsolete page observations from being reused as current state.
% At each iteration, the information update module first captures a new summary of the browser state, including a screenshot and a structured representation of the DOM. The screenshot is captured and used as runtime visual evidence at every execution step, regardless of whether the original bug report contains visual attachments. Therefore, \name’s use of visual evidence does not depend on whether the original bug report contains screenshots or other visual attachments. Report-level visual attachments are treated as optional input artifacts when available, whereas runtime screenshots are collected uniformly for all reports and are used for state grounding and final-state validation. During the \textbf{Browser-Info Update}, \name converts the raw browser snapshot into a compact browser-observable record, retaining only evidence that can ground later decisions, such as the visible DOM, interactable elements, URL, tab state, screenshot reference, and latest action outcome, and writes this record to the \textbf{Information Center} as the current state. Historical Filtering keeps the decision context compact by separating the latest-state view from the full trace. For each new step, the current browser state message replaces the previous state message, while the agent history is summarized and optionally truncated to the most relevant recent records. The complete action and browser-state trace is still saved for offline analysis, but obsolete page observations are not repeatedly exposed as current grounding evidence.

Based on the updated web information and the remaining plan, the agent decides the next execution step and produces one or more structured low-level actions that explicitly state the intended operation together with its required parameters. The supported browser actions include navigation, clicking, text input, keyboard input, scrolling, hovering, dropdown selection, file upload, waiting, and task termination. Each action is grounded in the current browser state and parameterized by fields such as an element index, URL, input value, file path, option text, or wait condition. The action specification is then handed to the web execution script, which materializes the operation in the browser through tool mediated interactions. After the action completes, its outcome is returned to the \textbf{Information Center} together with lightweight feedback, and the \textbf{Plan List Update} is performed to reflect the progress.
% , marking finished steps, revising or reordering remaining steps when the newly observed state suggests deviations from the original plan.

The Information Center is \name's structured execution-time memory. It stores the agent state, context summary, message history, action results, browser-state history, and reproduction statistics. Components read from and update this shared memory: the Context Builder records prepared artifacts, the Action Generator uses the latest browser state and action history, the Plan List Update module tracks the current plan step, and the Evaluator reads final-state evidence and outcome descriptions. This keeps subsequent actions grounded in the most recent observed state. 

The Action Generator emits structured JSON containing an assessment of the previous step, compact memory, the next immediate goal, the current high-level plan-step index, and one or more tool actions. The tool actions are validated against typed action schemas before execution, and actions can only refer to interactive element indices exposed in the current browser state.

% \paragraph{Tool-mediated actions with robustness guards.}
% To reduce brittleness, the executor performs robustness checks after each action.
% % Actions are executed through a guarded routine that enforces several runtime safety checks.
% The executor prevents invalid termination patterns by allowing task completion only when the model issues a single completion action for the step. It also detects page changes after each action by comparing element identities across successive DOM snapshots. If the target element has moved or if new interactive elements appear, remaining actions are aborted and control returns to the next decision step. This design improves robustness against dynamic GUIs and stale element references.

\textit{Failure control, budgets, and termination.}
The execution enforces explicit budgets, including a maximum number of execution steps, per action timeouts, and a limit on consecutive failures. An action is counted as failed when the executor cannot complete it because the target is unavailable, hidden, blocked, invalid, or the action times out. The consecutive-failure counter is reset after any successfully executed browser action. When failures accumulate beyond these thresholds, the execution forces a final step in which only termination is permitted. This design ensures that each run ends cleanly and produces a complete execution trace, even when reproduction cannot be completed, as detailed in \cref{subsec: experimental_setup}.

% \paragraph{Completion validation via final-state judging.}
% When a run terminates with completion, \name optionally performs automatic final-state evaluation. The executor compares the final browser state summary and the agent’s concluding description against report-derived specifications of expected and unexpected outcomes. An LLM-based judge determines whether the observed state satisfies the expected conditions while avoiding the unexpected ones. The resulting decision is aligned with the run’s final success status to ensure consistency across execution history and aggregated statistics.

\textit{Trace and statistics collection.}
Throughout execution, the system records screenshots, URLs, issued actions, execution outcomes, and step metadata into a structured trace as \textbf{Reproduction Report}. It also updates reproduction statistics online, including action execution success and coarse task progress inferred from step signals. These artifacts support downstream evaluation, error analysis, and ablation studies.

\subsubsection{Evaluation Phase}
After executing the report-described procedure, \name summarizes the final browser-observable state using a compact evidence representation. This summary includes the current URL, a structured view of the relevant visible elements and their states, screenshots, and the accumulated evidence stored in the \textbf{Information Center}. Based on the bug report, \name derives an outcome oracle that captures the reported failure symptom or the expected observable behavior, such as an error message, missing or incorrect UI content, unexpected navigation, or the inability to complete an operation. The evaluator checks the final evidence against this oracle and outputs a binary reproduction decision together with a short rationale grounded in the captured evidence. In implementation, the evaluator is an LLM-based judge that compares the final URL, visible DOM summary, screenshot evidence, and relevant trace entries with the expected and unexpected descriptions. This design allows practitioners to run \name against the target deployment under test and obtain a traceable reproduction result. The evaluator returns a structured decision indicating whether the final state satisfies the expected description, whether it satisfies the unexpected description, and an evidence-grounded rationale. A run is counted as successful only when it reaches the planned terminal state, satisfies the expected description, and does not satisfy the unexpected description.

%% file: 4-Study_design.tex
\section{Study Design}\label{sec:subject_study}

\subsection{Dataset}

\subsubsection{Criteria for Selecting Projects}

We select mature, actively maintained, and issue-rich web applications from diverse domains. Each selected project has more than 40,000 stars, 10,000 commits, and 4,000 issues, and was active within three months before data collection. The final subjects are Ghost, Metabase, NocoDB, and n8n, covering digital publishing, business intelligence, collaborative data platforms, and workflow automation.

\subsubsection{Dataset Collection}
Our goal is to evaluate whether a system can transform real-world GitHub bug reports into browser-executable reproduction procedures and check the resulting browser-observable states. We construct a new dataset because existing bug report corpora typically do not provide runnable web application deployments or report-derived outcome oracles, while web-agent benchmarks mainly focus on task completion rather than bug reproduction. We evaluate reports on controlled deployments of the selected applications rather than reconstructing a historical buggy snapshot for every report, because doing so at this scale would require identifying fixing changes, rebuilding prior versions and dependencies, and restoring compatible data and configuration. The dataset therefore focuses on reports with executable browser-level procedures and browser-observable end states, with annotations that support checking the final state on the evaluated deployment.

% We evaluate reproduction on controlled deployments of the selected applications rather than reconstructing a historical buggy snapshot for each report. Reconstructing hundreds of historical snapshots would require identifying fixing changes, rebuilding prior versions and dependencies, and restoring compatible data and configuration, which would substantially limit scalability. Therefore, our dataset focuses on reports that describe executable browser-level procedures and browser-observable end states. The original historical symptom may not always be observable in the evaluated version; accordingly, each report is annotated with outcome information that supports post hoc checking using browser-observable evidence.

We construct the dataset in two stages, as described below.

\noindent\textbf{Stage 1: bug report collection and validity confirmation.}
We collect issues from each project's GitHub issue tracker and identify confirmed bug reports. We first use an automated filter based on bug-related labels and simple text patterns in issue descriptions and discussions. We then manually review the candidates and remove false positives, such as feature requests mislabeled as bugs or discussions that do not describe a concrete defect. This process yields approximately 4,000 initial candidates and 875 confirmed bug reports, as shown in \cref{tab:dataset-stage-counts} (\#Stage 1 column).
For each confirmed report, we retain the original issue text and metadata needed for later execution and analysis. This procedure follows common practices for constructing bug report datasets from GitHub~\cite{bettenburg2008makes, zhou2012should, chaparro2017detecting, davies2014s}.

%We collect issues from each project's GitHub issue tracker and filter them into a set of confirmed bug reports. The objective is to determine whether an issue actually indicates a defect, rather than to evaluate the report's level of detail or quality. We start with an automated filter script that selects candidates using bug related labels and simple text patterns in the issue description and discussion. We then manually review the selected candidates and remove false positives, including feature requests mislabeled as bugs and discussions that do not describe a concrete problem. This two-step process ensures the scalability and accuracy of data collection, resulting in approximately 4,000 initial candidates and 875 confirmed bug reports, as shown in \cref{tab:dataset-stage-counts} (\#Stage 1 column).

% For each confirmed report, we maintain the original text and document the necessary information for subsequent execution and analysis. This includes the issue title and description, labels, comments, timestamps, and links to relevant pull requests or commits when available. Overall, this procedure follows common practices for building bug report datasets from GitHub, combining automated selection with focused manual checking to ensure reliable results while maintaining real-world report features~\cite{bettenburg2008makes, zhou2012should, chaparro2017detecting, davies2014s}.

\noindent\textbf{Stage 2: refinement and reproduction-oriented annotation.}
After Stage 1, many reports are still unsuitable for scalable browser-level evaluation. We remove reports that lack actionable user interactions or observable UI outcomes, while retaining reports that omit common but inferable steps, such as logging in or navigating to a relevant module. For each retained report, two authors annotate an expected description and an unexpected description. The expected description captures the final browser-observable state when the report-derived procedure is executed on the controlled evaluated deployment, whereas the unexpected description is derived from the failure symptom described in the original report. We further exclude reports matching one or more of the following conditions:

% The expected description is obtained by inspecting the behavior observed when the report-described procedure is executed on the controlled bug-absent deployment. It captures the intended final browser-observable state after the procedure is completed. In contrast, the unexpected description is derived from the failure symptom described in the original bug report. We further exclude reports matching one or more of the following conditions:

\begin{itemize}
  \item \textbf{(R1) Installation, build, or upgrade defects:}
  Reproduction primarily depends on dependency resolution, build tooling, or version upgrade workflows rather than web GUI interactions.

  \item \textbf{(R2) Specialized infrastructure:}
  Reproduction requires non-public resources such as proprietary databases, enterprise services, private integrations, or specialized hardware.

  \item \textbf{(R3) Version Locked issues:}
  The failure is explicitly tied to a past software version or historical configuration that cannot be recreated in the current deployed environment.

  \item \textbf{(R4) Non GUI failures:}
  Diagnosis or validation relies on backend logs, server traces, or other artifacts not accessible to a browser-based executor.
\end{itemize}

These filters exclude bug reports that are inherently incompatible with scalable browser execution, rather than reports that are merely challenging for our approach. After Stage 2, we obtained 667 bug reports across the four selected projects, as summarized in \cref{tab:dataset-stage-counts}. To reduce potential bias, two authors independently reviewed each report, and disagreements were resolved through discussion. As complementary validation, we construct a historical replay subset for RQ1 by sampling 10 successful \name runs per application and restoring the corresponding buggy versions when feasible.

% To complement the current-deployment evaluation, we additionally construct a small historical replay subset for RQ1. For each application, we first randomly sample 10 reports from successful \name runs in the main evaluation, before checking or restoring their historical buggy versions. We sample from successful runs because this replay study aims to check whether reconstructed procedures can expose the original bug-present behavior on restored buggy versions, rather than to estimate historical reproduction success over all reports.

\begin{table}[!t]
\centering
\small
\setlength{\tabcolsep}{5pt}
\begin{tabular}{lccc}
\toprule
\textbf{Projects} & \textbf{\#Stars}   & \textbf{\#Stage 1} & \textbf{\#Stage 2} \\
\midrule
Ghost    & 51.6k & 249 & 194 \\
Metabase & 45.6k & 203 & 157 \\
NocoDB   & 61.5k & 200 & 143 \\
n8n      & 169k  & 223 & 173 \\
\midrule
\textbf{Total} & -  & 875 & \textbf{667} \\
\bottomrule
\end{tabular}
\caption{Projects. Number of retained bug reports after Stage 1 and Stage 2 for each project.}
\label{tab:dataset-stage-counts}
\end{table}

\subsection{Research questions}
To systematically evaluate \name, we design the following research questions:

\noindent\textbf{RQ1. Effectiveness}: \textit{How effective is \name in reconstructing and executing report-described web GUI reproduction procedures from natural language bug reports?}

RQ1 evaluates the end-to-end effectiveness of \name on real bug reports. We treat effectiveness as executability: given a report, the system should interpret the described procedure, recover feasible missing prerequisites, produce a reproduction plan, and execute the required browser interactions. We run \name on the reports in our dataset while preserving their noise and incompleteness, and provide only the minimal setup needed for controlled execution.

We compare \name with two external baselines and multiple LLM backbones, and report RSR, TCR, and AESR from validation outputs and execution traces. We further perform a project-balanced historical replay on successful runs to assess whether reconstructed procedures can trigger the original reported failures on restored buggy versions.

% We compare \name against two external baselines that represent common alternatives for automated web reproduction: \emph{Direct script generation}, which prompts an LLM to synthesize a runnable browser automation script from the report in one shot, and \emph{Browser-use}, a browser agent baseline that interacts with a real browser. We also run \name with different LLM backbones under the same experimental setting to examine model sensitivity. The pipeline and environment remain fixed, and only the model used for planning, decision making, and validation prompts varies. We report reproduction success rate, task completion rate, and action execution success rate derived from validation outputs and execution traces. For a project-balanced subset of successful runs, we further replay the corresponding reports on historical buggy versions to examine whether the reconstructed procedures can trigger the original reported failures.

\vspace{0.5em}
\noindent\textbf{RQ2. Ablation}: \textit{How do context reconstruction and plan generation affect the effectiveness of automated web GUI bug reproduction?}

RQ2 examines the role of two preparation-stage components. First, we disable the Context Builder while keeping the deployed instances fixed and the success criteria unchanged. Many bug reports depend on prerequisites that may be unavailable at execution time; by disabling this module and executing the raw report with only the minimal environment setup, we quantify the contribution of explicit prerequisite reconstruction. Second, we disable the Plan Generator and let the execution agent proceed directly from the bug report and available context. This variant examines whether the initial high-level intent scaffold contributes beyond the online planning and state-aware action selection already performed during execution.

\vspace{0.5em}
\noindent\textbf{RQ3. Failure Analysis}: \textit{Why does the reproduction fail in practice, and what are the dominant failures across different web applications?}

% RQ3 investigates why reproduction runs fail and which failure modes are prevalent across applications. Unlike RQ1 and RQ2, which measure effectiveness and isolate design contributions, this question focuses on diagnosing the bottlenecks that prevent successful execution. The objective is to identify the obstacles that most frequently halt progress so that future improvements toward the factors that truly limit reproduction.

% We perform a retrospective analysis of failed runs. For each report, the system generates an execution trace that records observations, issued actions, action outcomes, plan progress, and a final pass flag that indicates success or failure, optionally with a short failure note. We inspect the traces of unsuccessful runs and assign one or more failure labels using a shared taxonomy. In most cases, a run has a single dominant label, and we allow multiple labels when distinct blockers are observed.

RQ3 investigates the bottlenecks that prevent successful reproduction. We perform a retrospective analysis of failed runs using execution traces that record observations, issued actions, action outcomes, plan progress, and the final pass flag with an optional failure note. We inspect unsuccessful runs and assign one or more failure labels using a shared taxonomy. In most cases, a run has a single dominant label, while multiple labels are used when distinct blockers are observed.

\subsection{Evaluation Metrics} \label{sec:metrics}
We evaluate \name using the following metrics that capture end-to-end reproducibility, task level progress, and action-level executability. Let $D = \{r_i\}_{i=1}^{N}$ denote a dataset of $N$ bug reports. We execute each report for multiple independent attempts and compute the metrics for each attempt. In Eqs.~\eqref{eq:rsr}, \eqref{eq: eq2}, \eqref{eq: eq3}, and \eqref{eq:failprop}, each attempt is treated as a run level sample. Accordingly, $N$ denotes the total number of runs across all reports. Each run produces an execution trace that records issued actions, plan status, and observed browser states, together with a final user observable state $s_i$ captured at termination.

\subsubsection{Reproduction Success Rate (RSR)}
In our evaluation setting, the buggy symptom described in a report may not be observable on the evaluated deployment because the reports are closed and we do not reconstruct a historical buggy snapshot per issue. We therefore adopt a metamorphic testing inspired oracle. Under this setting, reproduction success is interpreted as end-to-end procedural completion with end state consistency under the annotated expected and unexpected oracle pair. Each report $r_i$ is annotated with an expected state description $E_i$, representing the intended or bug absent behavior, and an unexpected state description $U_i$, representing the erroneous or bug present behavior. Given the final state $s_i$, let $m_E(i)\in\{0,1\}$ and $m_U(i)\in\{0,1\}$ indicate whether $s_i$ matches $E_i$ and $U_i$, respectively. A run is counted as a successful reproduction if it terminates as completed ($\mathrm{done}_i=1$), matches the expected state, and does not match the unexpected state. We define:
\begin{equation}
\mathrm{RSR}:=\frac{1}{N}\sum_{i=1}^{N}
\mathbb{I}[\mathrm{done}_i \wedge m_E(i) \wedge \neg m_U(i)] .
\label{eq:rsr}
\end{equation}

The variable $\mathrm{done}_i$ indicates that the agent terminated with the task marked as completed, meaning that it reached the final planned goal state. This enforces the rule that only runs that complete the task are eligible to be counted as successful reproductions. The key idea of RSR is consistency with the annotated expected and unexpected oracle pair. For the historical replay subset, we separately report whether the restored buggy version exhibits the report-described bug-present behavior after executing the reconstructed procedure. This metric is not used to replace RSR; instead, it provides complementary evidence that successful current-deployment procedures can correspond to actual failure-triggering executions when the relevant buggy version is restored. We compute RSR at the run level. Each report is executed for multiple independent attempts under the same configuration, and each attempt contributes one run-level sample to Eq.~\eqref{eq:rsr}. Therefore, the reported RSR values are aggregated over all attempts across reports, rather than computed from a single execution per report.

\subsubsection{Task Completion Rate (TCR)}
% \textbf{Design rationale.}
RSR is intentionally strict and binary, and therefore does not reflect partial progress toward reproduction. TCR captures how far the system progresses, even when it does not fully complete the task. For each report $r_i$, let $P_i$ be a \emph{flat} sequence of $|P_i|$ high-level plan steps produced during preparation, and let $C_i \subseteq \{1,\dots,|P_i|\}$ be the indices of plan steps marked as completed in the plan-status trace. We define:
\begin{equation}
\mathrm{TCR}(i) := \frac{|C_i|}{|P_i|} \in [0,1],
\label{eq: eq2}
\end{equation}

% $P_i$ is a linear sequence of high-level guidance steps produced during preparation, and $C_i$ directly corresponds to the set of completed step indices inferred from the plan-status trace. In our results, we report the mean Task Completion Rate and additionally recommend reporting the median and interquartile range to better reflect skewed completion behavior.

\subsubsection{Action Execution Success Rate (AESR)}
% \textbf{Design rationale.}
AESR measures action level executability by quantifying whether issued actions can be successfully executed by the automation backend. It complements RSR and TCR because an agent may fail individual actions during exploration and still recover to complete the task, and it measures the precision of action execution. For report $r_i$, let $A_i$ be the sequence of issued actions with length $|A_i|$, and let $x_{i,j}\in\{0,1\}$ denote whether the $j$-th action is executed successfully. We define:
\begin{equation}
\mathrm{AESR}(i) := \frac{1}{|A_i|} \sum_{j=1}^{|A_i|} x_{i,j},
\label{eq: eq3}
\end{equation}

\subsubsection{Cost metrics}
We compute interaction and time level cost from the same execution traces. For each report $r_i$, we record (i) the number of issued actions $n_{\mathrm{act}}(i):=|A_i|$; (ii) wall-clock duration $T(i):=t_i^{\mathrm{end}}-t_i^{\mathrm{start}}$; (iii)and the number of LLM calls $n_{\mathrm{llm}}(i):=\sum_{j=1}^{|A_i|}\mathbb{I}\!\left[\mathrm{llmcall}(a_{i,j})=1\right]$.

% \begin{equation}
% n_{\mathrm{act}}(i) := |A_i|, \quad
% n_{\mathrm{fail}}(i) := \sum_{j=1}^{|A_i|} \mathbb{I}[x_{i,j}=0]
% \end{equation}

% \begin{equation}
% n_{\mathrm{retry}}(i) := \text{number of retries or backtracking steps inferred from the trace}
% \end{equation}

% \begin{equation}
% T(i) := t^{\mathrm{end}}_i - t^{\mathrm{start}}_i
% \end{equation}

% In the paper, we report the distributions of interaction and time cost using the median and interquartile range for $n_{\mathrm{act}}$, $n_{\mathrm{fail}}$, $n_{\mathrm{retry}}$, and $T$.

\subsubsection{Failure taxonomy distribution}
% We summarize failures at the level of \emph{failure labels} rather than unique failed runs, since a single run can expose multiple issues during execution and therefore contribute to multiple categories. For each report run $i$, we extract a (possibly empty) multiset of failure labels $\mathcal{L}_i$ from the reproduction summary and the execution log analysis. Each label $\ell \in \mathcal{L}_i$ is mapped to one taxonomy category in $\mathcal{C}$, but a run may yield multiple labels that map to different categories.

% Let $\mathcal{I}_p$ denote the set of runs associated with project $p$. We define the total number of collected failure labels in project $p$ as

We summarize failures at the level of \emph{failure labels} rather than unique failed runs. For each run $i$, we extract a multiset of failure labels $\mathcal{L}_i$, where each label $\ell\in\mathcal{L}_i$ maps to exactly one taxonomy category in $\mathcal{C}$. Let $\mathcal{I}_p$ be the set of runs associated with project $p$ and let $N^{\mathrm{lab}}_p := \sum_{i\in\mathcal{I}_p}|\mathcal{L}_i|$ be the total number of collected labels. For a category $k\in\mathcal{C}$, we compute:
\begin{equation}
\mathrm{FailProp}_p(k) :=
\frac{\sum_{i \in \mathcal{I}_p} \sum_{\ell \in \mathcal{L}_i} \mathbb{I}[\mathrm{cat}(\ell)=k]}
{N^{\mathrm{lab}}_p}.
\label{eq:failprop}
\end{equation}

\subsection{Baselines}
We compare \name with two external baselines and report model variants of \name to quantify backbone sensitivity.

\textit{\textbf{Backbone variants of \name.}}
We run \name with different backbone LLMs, including GPT 5 mini, Claude 4.5 Haiku, and Gemini 2.5 Flash. This comparison keeps the system design and environment fixed and varies only the underlying model.

\textit{\textbf{Browser-use.}}
Browser-use is a representative web agent that interacts with a real browser without our preparation stage and workflow stabilizations\footnote{Browser-use: the AI browser agent, \url{https://github.com/browser-use/browser-use}}.

\textit{\textbf{Direct script generation.}}
This baseline prompts an LLM to synthesize a runnable Playwright or Selenium script from the bug report in one shot. We instantiate it with GPT 5 mini and Gemini 2.5 Flash.

\subsection{Experimental Setup}\label{subsec: experimental_setup}

\textit{\textbf{Target applications and deployment.}}
We evaluate the system on four widely used open source web applications: Ghost, Metabase, NocoDB, and n8n. Each application is deployed using its officially recommended self-hosting method to keep the environment realistic and reproducible. Ghost is installed locally using Ghost-CLI in development mode with an SQLite database. Metabase is deployed using its official Docker image with a persistent application database following its production guidance. NocoDB is deployed using Docker Compose to run the application together with its database service. n8n is deployed using its official Docker workflow with persistent storage. All methods are evaluated on the same deployed instances.

% We evaluate \name against two external baselines that represent common alternatives for automated web reproduction. Direct script generation prompts an LLM to synthesize a runnable Playwright or Selenium script from the bug report in one shot, instantiated with GPT 5 mini and Gemini 2.5 Flash. Browser-use serves as a browser agent baseline that drives a real browser without our preparation stage and workflow level stabilizations. In addition, we report \name with three different backbone LLMs as model variants to quantify sensitivity under the same system design.

\textit{\textbf{Browser execution environment.}}
All reproductions are executed in a controlled browser automation environment. We use a stable Chrome-based browser with a fixed viewport and consistent language and locale settings to reduce UI variation across runs. During execution, we collect browser-observable signals required for validation, including DOM snapshots, screenshots, and element-level state information. All configurations, including the script generation baseline, run on the same deployed application instances and under the same browser environment.

\textit{\textbf{Global Parameter Settings.}}
To ensure comparability across research questions and ablation studies, we keep the following parameters fixed unless explicitly stated otherwise.

\begin{itemize}
  \item \textit{Execution budgets.} For each report, we enforce a maximum of $B_{\mathrm{step}}=20$ executed steps and cap the number of allowed execution steps failures at $B_{\mathrm{fail}}=3$. In addition, we apply a per action execution time limit of $T_{\mathrm{act}}=120\text{s}$ to bound the runtime of individual automation steps and avoid indefinite blocking on slow or unstable page states.
  
  \item \textit{Termination criteria.} A run terminates when the agent declares task completion, when the validation procedure ends the run, or when the execution budgets are exceeded.
  
  \item \textit{State initialization.} For each application, we standardize the initial execution state by using fixed accounts or credentials and a clean baseline configuration. This ensures that all reports are executed from comparable starting points across runs. For each sampled historical replay case, we identify the relevant buggy version from issue, pull request, commit, or release information, and deploy that version when feasible. The replay uses the same report-driven \name workflow and browser execution environment as the main evaluation, but the final decision checks for the bug-present behavior described in the original report.

  \item \textbf{Repeated attempts and aggregation.} Each report is executed for up to $K=3$ independent attempts. We use an early-stop protocol: once a report is successfully reproduced, subsequent attempts for that report are skipped. As a result, the actual number of executions for a dataset with $N$ reports ranges from $N$ to $3N$. RSR, TCR, AESR, action count, time, and LLM calls are computed over the executed attempts and reported as per-run means.
\end{itemize}

% \paragraph{Execution budgets.}
% For each report, we enforce a maximum wall-clock execution time $T_{\max}$ and a maximum number of executed actions $B_{\mathrm{act}}$. These budgets bound execution cost and prevent pathological behaviors such as unbounded exploration loops.

% \paragraph{Termination criteria.}
% A run terminates when the agent declares task completion, when the validation procedure ends the run, or when the execution budgets are exceeded.

% \paragraph{Model inference settings.}
% Decoding parameters and inference-related settings are held constant across all experimental conditions, including system prompts and tool schemas. In ablation studies, we vary only the backbone model while keeping all other inference parameters unchanged.

% \paragraph{State initialization.}
% For each application, we standardize the initial execution state by using fixed accounts or credentials and a clean baseline configuration. This ensures that all reports are executed from comparable starting points across runs.

\subsection{Reproducibility artifacts}\label{subsec:reproducibility_artifacts} To support reproducibility, we provide an \href{https://zenodo.org/records/18417124?preview=1&token=eyJhbGciOiJIUzUxMiJ9.eyJpZCI6IjA2NzI5MTczLTJhMmMtNDlmOS04ODAyLTZjODU0NzlkODcyMiIsImRhdGEiOnt9LCJyYW5kb20iOiI0MDk3ZDNjMjk4OTk1ZDcwNWVkZDFiODI3ODNjOGZhYSJ9.-evKy-AeflcQTm50OAjByOSdHKn7WULPp1_Av2tJDZ_BYTKPQvJFXd_biVGLyHSE_yRhlh_qpWXtCIX5qqpcJQ}{\textcolor{blue}{\uline{anonymous artifact repository}}} containing the implementation, prompt templates, structured output schemas, evaluator rubric, tool specifications, model settings, execution configuration, and representative traces needed to inspect and reproduce the experiments.

% containing the implementation, prompt templates, structured output schemas, evaluator rubric, tool specifications, model settings, execution configuration, and representative reproduction traces. In the paper, we summarize the key design choices needed to understand the approach, while the repository provides the full executable artifacts and detailed prompt-level specifications.

%% file: 5-Results.tex
\section{Results}\label{sec:results}

\subsection{\textbf{RQ1. Effectiveness}}\label{sec: result_rq1}
% Tables~\ref{tab:rq1-baselines} and~\ref{tab:rq1-main} summarize RQ1 results. Table~\ref{tab:rq1-baselines} compares \name with external baselines using reproduction success rate (RSR). Table~\ref{tab:rq1-main} reports effectiveness and cost for \name under different backbone models.
To contextualize effectiveness, we compare \name with two external baselines. Direct script generation is a simple baseline that asks an LLM to synthesize a runnable Playwright or Selenium script from the report in one attempt. Browser-use represents a browser agent baseline that interacts with a real browser without our preparation stage, enhanced tool suite, and execution safeguards. In addition, we run \name with different backbone models to quantify model sensitivity under the same system design. Since the baselines do not provide compatible traces for fine-grained metrics, we report the reproduction success rate (i.e., \textit{RSR}) for the method comparison and use the full set of metrics and cost statistics when comparing LLM backbone variants of \name.

\begin{table}[t]
\centering
\footnotesize
\setlength{\tabcolsep}{3pt}
\renewcommand{\arraystretch}{1.06}
\begin{tabularx}{\columnwidth}{@{}lXrr@{}}
\toprule
\textbf{Dataset} & \textbf{Approach} & \textbf{RSR} & \textbf{$\Delta$RSR} \\
\midrule
\multirow{4}{*}{Ghost}
& \textbf{\name (Full), GPT-5 mini} & \textbf{45.79} & — \\
& Browser-use, GPT-5 mini & 30.92 & -14.87 \\
& Direct script gen., GPT-5 mini & 4.00 & -41.79 \\
& Direct script gen., Gemini-2.5 Flash & 4.82 & -40.97 \\
\midrule

\multirow{4}{*}{Metabase}
& \textbf{\name (Full), GPT-5 mini} & \textbf{51.75} & — \\
& Browser-use, GPT-5 mini & 40.63 & -11.12 \\
& Direct script gen., GPT-5 mini & 3.80 & -47.95 \\
& Direct script gen., Gemini-2.5 Flash & 3.10 & -48.65 \\
\midrule

\multirow{4}{*}{NocoDB}
& \textbf{\name (Full), GPT-5 mini} & \textbf{49.68} & — \\
& Browser-use, GPT-5 mini & 39.47 & -10.21 \\
& Direct script gen., GPT-5 mini & 2.76 & -46.92 \\
& Direct script gen., Gemini-2.5 Flash & 2.76 & -46.92 \\
\midrule

\multirow{4}{*}{n8n}
& \textbf{\name (Full), GPT-5 mini} & \textbf{52.60} & — \\
& Browser-use, GPT-5 mini & 35.29 & -17.31 \\
& Direct script gen., GPT-5 mini & 1.34 & -51.26 \\
& Direct script gen., Gemini-2.5 Flash & 2.24 & -50.36 \\
\midrule

Average & \textbf{\name (Full), GPT-5 mini} & \textbf{49.96} & — \\
\bottomrule
\end{tabularx}
% \caption{RQ1 method comparison using reproduction success rate (RSR).}
\caption{RQ1 method comparison using reproduction success rate (RSR). $\Delta$RSR denotes the absolute decrease in RSR (\%) relative to the full \name configuration with GPT 5 mini on the same dataset. Negative values indicate lower performance compared to \name (Full). We include two external baselines: (i) \textbf{Direct script generation}, a one-shot baseline that asks an LLM to synthesize a runnable Playwright or Selenium script from the report, and (ii) \textbf{Browser-use}, a browser agent baseline that drives a real browser without our preparation stage, enhanced tool suite, and execution safeguards.}
\label{tab:rq1-baselines}
\end{table}

Table~\ref{tab:rq1-baselines} summarizes the reproduction success rates (RSR) of \name and the baselines. 
The full \name configuration with GPT 5 mini consistently achieves the highest RSR, ranging from 45.79\% (Ghost) to 52.60\% (n8n). The \textit{Browser-use} baseline achieves moderate success (30.92\%–40.63\%) when it runs a real browser without any setup or workflow-level stabilizations. Its performance consistently falls between 10\% and 17\% compared with \name (Full), emphasizing the significance of structured execution and context reconstruction. \textit{Direct script generation}, a one-shot LLM baseline, performs poorly across all datasets (1.34\%–4.82\%), showing that reproducing real-world bug reports requires more than translating natural language into scripts. Representative failures show that one-shot script generation often fails to handle missing setup and dynamic UI states, and Browser-use can interact with live pages, but it remains less reliable for bug reproduction because it lacks report-specific context reconstruction, artifact preparation, and reproduction-oriented safeguards for blocked or incomplete execution states.

\begin{table}[!htbp]
\centering
\small
\setlength{\tabcolsep}{3.5pt}
\renewcommand{\arraystretch}{1.1}
\resizebox{\columnwidth}{!}{%
\begin{tabular}{ll|ccc|ccc}
\toprule
\multirow{2}{*}{\textbf{Dataset}} &
\multirow{2}{*}{\textbf{LLM Model}} &
\multicolumn{3}{c|}{\cellcolor{LightGray}\textbf{Effectiveness Metrics}} &
\multicolumn{3}{c}{\cellcolor{LightGray}\textbf{Cost Metrics}} \\
\cmidrule(lr){3-5}\cmidrule(lr){6-8}
& & \textbf{RSR \%} & \textbf{TCR \%} & \textbf{AESR \%}
& \textbf{\#Actions} & \textbf{Time} & \textbf{\#LLM Calls} \\
\midrule

\multirow{3}{*}{Ghost}
& GPT 5 mini         & 45.79 & 72.16 & 82.93 & 32.23 & 7.27 & 16.42 \\
& Claude 4.5 Haiku   & 49.48 & 75.83 & 86.41 & 31.67 & 7.11 & 16.37 \\
& Gemini 2.5 Flash   & 47.94 & 73.02 & 84.17 & 29.81 & 5.97 & 14.19 \\
\midrule

\multirow{3}{*}{Metabase}
& GPT 5 mini         & 51.75 & 77.29 & 81.57 & 27.68 & 7.70 & 18.90 \\
& Claude 4.5 Haiku   & 50.96 & 75.11 & 79.02 & 28.39 & 7.53 & 18.27 \\
& Gemini 2.5 Flash   & 51.59 & 76.88 & 80.41 & 26.21 & 6.41 & 16.73 \\
\midrule

\multirow{3}{*}{NocoDB}
& GPT 5 mini         & 49.68 & 76.87 & 92.46 & 34.10 & 7.74 & 19.05 \\
& Claude 4.5 Haiku   & 53.15 & 74.23 & 90.11 & 33.77 & 7.21 & 18.79 \\
& Gemini 2.5 Flash   & 54.55 & 78.94 & 91.33 & 31.93 & 6.19 & 16.61 \\
\midrule

\multirow{3}{*}{n8n}
& GPT 5 mini         & 52.60 & 73.58 & 89.18 & 27.92 & 7.25 & 16.56 \\
& Claude 4.5 Haiku   & 52.02 & 71.37 & 90.47 & 28.63 & 6.97 & 16.19 \\
& Gemini 2.5 Flash   & 51.45 & 74.91 & 88.21 & 25.97 & 5.81 & 13.97 \\
\bottomrule
\end{tabular}%
}
\caption{RQ1 effectiveness and cost of \name with different LLM models across four web applications, averaged over multiple independent runs per report. \textbf{RSR\%} measures end-to-end reproduction success. \textbf{TCR\%} denotes the proportion of planned steps completed per run, and \textbf{AESR\%} denotes the proportion of actions executed successfully. \textbf{\#Actions}, \textbf{Time} (minutes), and \textbf{\#LLM Calls} denote per-run means.}
\label{tab:rq1-main}
\end{table}

% Table~\ref{tab:rq1-main} further presents the effectiveness and cost metrics of \name when paired with three different LLM models across our dataset. The metrics capture reproduction success rate (\emph{RSR}), task completion rate (\emph{TCR}), action execution success rate (\emph{AESR}), and computational cost.
% Across all applications and LLM models, \name achieves consistent end-to-end reproduction success: \emph{RSR} ranges from 45.79\% to 54.55\% depending on the dataset and backbone model, indicating that the system can reproduce a substantial portion of real bug reports in a fully executable setting. Beyond strict success, \name also makes strong partial progress.

Table~\ref{tab:rq1-main} further reports the effectiveness and cost of \name with three LLM backbones. Across all applications and models, \name shows stable end-to-end performance: RSR ranges from 45.79\% to 54.55\%, TCR stays around the mid-70\% range, and AESR remains high across most settings. These results indicate that many runs complete most planned steps and execute most low-level actions successfully, even when they do not satisfy the strict final reproduction oracle. The gap between high action executability and lower final RSR suggests that a small number of brittle interactions or late-stage mismatches can still determine whether the strict oracle is satisfied.

% Specifically, although n8n achieves the highest \emph{RSR}, its canvas-centric editing with frequent modals and nested menus makes action grounding and execution less reliable. When reports provide an importable prebuilt workflow the \emph{RSR} rises, but building from scratch accumulates errors and reduces completion and action success. NocoDB and Metabase attain high \emph{TCR} and \emph{AESR} because reproductions consist of many small, straightforward UI actions that execute reliably. Yet the overall UI complexity and long dependency chains mean that missing prerequisites, unobserved state changes, or late-stage mismatches can still block reaching the final oracle, suppressing \emph{RSR}.\CHECK{...}

The project-level patterns also explain the remaining variation. n8n achieves the highest RSR, especially when reports provide importable workflows, but its canvas-centric editor, frequent modals, and nested menus make action grounding less reliable when workflows must be built from scratch. NocoDB and Metabase show high TCR and AESR because many reproductions consist of smaller, more direct UI actions; however, longer dependency chains, missing prerequisites, unobserved state changes, or late-stage mismatches can still prevent the final oracle from being satisfied.

% The cost metrics are moderate and stable. Successful and partially successful runs typically require on the order of 26–34 actions on average, complete within about 6–8 minutes on average, and invoke the LLM about 14–19 times on average. These costs are small relative to manual script authoring and debugging, which often involves iterative implementation, repeated reruns, and human troubleshooting when selectors or prerequisites break. 

% Replacing the backbone LLM produces some variation in results, but differences remain modest. Across the four datasets, the three models achieve similar \emph{RSR}. These observations indicate that the execution pipeline has a stronger influence on end-to-end executability than the choice of a single model.

The cost metrics remain moderate and stable. Runs require about 26-34 actions, 6-8 minutes, and 14-19 LLM calls on average. These costs are small compared with manual script authoring and debugging, which often require iterative implementation, repeated reruns, and troubleshooting when selectors or prerequisites break. Replacing the backbone LLM changes the numbers modestly, but the overall trends remain consistent, suggesting that the execution pipeline has a stronger influence on executability than the specific model choice.

% \paragraph{Historical buggy-version replay.}
\subsubsection{Historical buggy-version replay.}
The main RQ1 evaluation uses controlled current deployments, which enables large-scale comparison but cannot by itself show whether reconstructed procedures trigger the original failures on historical buggy versions. We therefore conduct a project-balanced historical replay study on 40 reports by randomly selecting 10 successful \name runs per application. This subset is not used to re-estimate overall success; instead, it checks whether procedures that pass the current-deployment oracle can expose the bug-present behavior after the corresponding buggy version is restored.

% The main RQ1 evaluation above is conducted on controlled current deployments, which supports large-scale and consistent comparison across methods, but does not by itself show whether the reconstructed procedures can trigger the original reported failures on historical buggy versions. To address this concern, we further perform a historical buggy-version replay study on a project-balanced sample of 40 reports. For each application, we randomly select 10 reports from the successful \name runs in the current-deployment evaluation. We select from successful runs because the goal of this additional study is not to re-estimate the overall success rate, but to validate whether procedures that pass the current-deployment oracle can also expose the original bug-present behavior when the corresponding buggy version is restored.

For each selected report, we identify the corresponding historical buggy version, rebuild and deploy that version when feasible, and rerun the same report-driven reproduction process against the restored deployment. A historical replay is counted as successful when the execution reaches the report-described scenario and the final browser-observable state matches the bug-present behavior described in the original report. ~\cref{tab:rq1-historical-replay} summarizes the results.

\begin{table}[t]
\centering
\setlength{\tabcolsep}{3.5pt}
\begin{tabular}{@{}lrrrr@{}}
\toprule
\textbf{Dataset} & \textbf{Sampled} & \textbf{Rebuilt} & \textbf{Triggered} & \textbf{Trigger Rate} \\
\midrule
Ghost    & 10 & 10 & 9  & 90.0\%  \\
Metabase & 10 & 10 & 10 & 100.0\% \\
NocoDB   & 10 & 10 & 10 & 100.0\% \\
n8n      & 10 & 10 & 10 & 100.0\% \\
\midrule
\textbf{Total} & 40 & 40 & 39 & 97.5\% \\
\bottomrule
\end{tabular}
\caption{Historical buggy-version replay on a project-balanced sample of current-deployment successful reports. \textbf{Sampled} denotes reports randomly selected from successful \name runs in the current-deployment evaluation. \textbf{Built} denotes historical buggy versions that were successfully reconstructed and deployed. \textbf{Triggered} counts reports for which \name triggered the report-described bug-present behavior on the corresponding buggy version.}
\label{tab:rq1-historical-replay}
\end{table}

As shown in Table~\ref{tab:rq1-historical-replay}, \name triggers the reported bug-present behavior in 39 out of 40 historical buggy-version replays. For all 40 replay cases, two authors manually inspected the final screenshots, DOM summaries, execution traces, and the report-described bug-present behavior. The manual judgments were consistent with the automated evaluator decisions in all cases, providing additional confidence in the reported replay outcomes. These results validate the semantic meaning of successful current-deployment runs: the reconstructed procedures usually expose the original failure once the corresponding buggy version is restored, rather than merely completing unrelated current-version workflows. The only unsuccessful Ghost case required a specific theme that was not materialized in the replay setup, so the symptom could not be observed even on the historical buggy version.

\begin{tcolorbox}[boxsep=2pt,left=2pt,right=2pt,top=1pt,bottom=1pt, before skip=10pt, after skip=5pt]
\noindent\textbf{Answer to RQ1.} 
\name reconstructs and validates a substantial portion of real-world web GUI bug-report procedures on controlled current deployments, achieving on average 49.96\% RSR and outperforming both the browser-agent baseline and one-shot script generation. The performance varies little across backbone LLMs, indicating that agent design influences reproducibility more than model choice. Historical replay further provides complementary evidence that procedures passing the current-deployment oracle can usually trigger the original report-described failures when the corresponding buggy versions are restored.
\end{tcolorbox}

\subsection{\textbf{RQ2. Ablation on System Components}}\label{sec: result_rq2}
Table~\ref{tab:rq2-context-ablation} shows how two preparation-stage components affect reproduction success. To keep the comparison focused, all RQ2 variants use the same deployed applications, success criterion, model backbone, and execution budgets as the full system. We report \emph{RSR} as the main metric for component-level comparison. 
% For the Plan Generator ablation, we remove the initial high-level plan and start browser execution directly from the bug report and available context, while keeping the same reports, model, environment, and execution budgets as in \name. This results in only a small decrease in RSR, suggesting that the initial plan mainly helps stabilize progress tracking, whereas most execution capability comes from the online state-aware action generation loop.

Removing the \emph{Context Builder} leads to a clear decrease in success on all evaluated projects. \emph{RSR} drops by 6.63\% to 11.29\%, with Ghost and n8n showing larger declines. This reduction indicates that many bug reports implicitly rely on missing prerequisites, such as entry information, required resources, or setup steps that are difficult to infer purely during online execution. Explicit materialization before browser interaction reduces early breakdowns and improves the chance of completing long reproduction procedures end-to-end.

Removing the \emph{Plan Generator} leads to only a small RSR decrease in our preliminary results. This is expected because the Plan Generator mainly provides a high-level intent scaffold and a reference for tracking progress, while the execution agent still observes the live browser state and plans the next action online through the Action Generator. Therefore, the result suggests that the initial plan is useful for structuring execution and progress monitoring, but the main reproduction capability comes from state-aware online execution and prerequisite reconstruction rather than from a fixed upfront plan alone.

Although a significant proportion of reports require additional context, the \emph{RSR} decrease after removing context reconstruction is smaller than that proportion. This is because some reports remain difficult even with reconstructed context, so disabling the module does not reduce success proportionally. In addition, certain prerequisites can still be fulfilled through online interaction, such as registering an account or preparing data within the application, which partially compensates for missing offline preparation.

\begin{table}[!t]
\centering
\setlength{\tabcolsep}{4pt}
\renewcommand{\arraystretch}{1.2}
\resizebox{\columnwidth}{!}{%
\begin{tabular}{llrrrr}
\toprule
\textbf{Condition} & \textbf{Setting} & \textbf{Ghost} & \textbf{Metabase} & \textbf{NocoDB} & \textbf{n8n} \\
\specialrule{1.2pt}{1.5pt}{1.5pt}
\rowcolor{RowGray}
Default
& \name, Full (\%)
& 45.79
& 51.75
& 49.68
& 52.60 \\
\midrule

Context ablation
& \name,
& 34.50 
& 44.68 
& 43.05 
& 42.20  \\
& w/o Context Builder (\%)
& \textbf{-11.29} & \textbf{-7.07} & \textbf{-6.63} & \textbf{-10.40} \\
\specialrule{1.2pt}{1.5pt}{1.5pt}

Plan ablation 
& \name, 
& 45.34
& 50.32
& 48.95
& 50.29 \\ 
& w/o Plan Generator (\%) 
& \textbf{-0.45} & \textbf{-1.43} & \textbf{-0.73} & \textbf{-2.31} \\ 
\specialrule{1.2pt}{1.5pt}{1.5pt}

Context requirement
& Req. extra context share (\%)
& 51.03
& 29.30
& 30.07
& 32.95 \\
\bottomrule
\end{tabular}}
\caption{RQ2 ablation results on the impact of Context Builder and Plan Generator on reproduction success rate (RSR) across projects. Values in parentheses indicate the absolute decrease in RSR (percentage points) relative to the full \name configuration. The last row reports the proportion of bug reports that require additional context preparation.}
\label{tab:rq2-context-ablation}
\end{table}

\begin{tcolorbox}[boxsep=2pt,left=2pt,right=2pt,top=1pt,bottom=1pt]
\noindent\textbf{Answer to RQ2.} 
Context Builder has a clear effect, while removing the Plan Generator causes only a small decrease, suggesting that the initial plan mainly stabilizes progress tracking and that online state-aware execution contributes most of the execution capability.
\end{tcolorbox}

\subsection{\textbf{RQ3. Reproduction Failure Analysis}}\label{sec: result_rq3} 

% To understand why reproduction attempts fail, we analyze unsuccessful runs and categorize their failure causes at two levels of granularity. Each failure is first assigned a fine-grained label based on execution traces, tool logs, and intermediate screenshots, and then grouped into four high-level categories: UI interaction failures, execution control failures, missing preconditions and dependency failures, and specification issues. 

To understand why reproduction attempts fail, we analyze unsuccessful runs using execution traces, tool logs, and intermediate screenshots. Each failed run is assigned one or more fine-grained failure labels and then mapped to broader failure sources: UI interaction failures, execution-control failures, missing preconditions or dependencies, and specification issues. To reduce subjectivity, two authors independently labeled the failures using predefined criteria and resolved disagreements through discussion.

Specifically, we label failures using a trace-guided coding protocol driven by two complementary evidence sources. First, we consult the reproduction summary produced by the system at the end of each run, which often includes an inferred failure explanation. Second, we inspect the execution logs to validate or revise this hypothesis by analyzing the full trajectory, including the last successfully executed actions, observed page transitions, and the final reachable browser state. When the summary and logs conflict, we prioritize the explanation supported by the logs. Failures with missing special annotations, namely $pass = N$ with no failure note, are conservatively assigned to \emph{Element grounding or interaction blocked}.

The taxonomy contains eight fine-grained labels: \emph{Element grounding or interaction blocked}, \emph{Unsupported primitive}, \emph{File/artifact prerequisite}, \emph{Data/database prerequisite}, \emph{Configuration/environment prerequisite}, \emph{Deployment/external dependency}, \emph{Report ambiguity or insufficient detail}, and \emph{Budget exceeded}. These labels are grouped into broader failure sources covering UI interaction failures, execution-control failures, missing preconditions or dependencies, and specification issues.

\begin{table}[t]
\centering
\footnotesize
\setlength{\tabcolsep}{3pt}
\renewcommand{\arraystretch}{1.05}
\begin{tabularx}{\columnwidth}{@{}Xcccc@{}}
\toprule
\textbf{Failure sub-category} & \textbf{Ghost} & \textbf{Metabase} & \textbf{NocoDB} & \textbf{n8n} \\
\midrule

\textbf{Element grounding / blocked} & 72.03 & 60.47 & 64.29 & 73.91 \\
Unsupported primitive & 1.69 & 2.33 & 2.38 & 3.26 \\

\midrule
File/artifact prerequisite & 6.78 & 6.98 & 8.33 & 5.43 \\
Data/database prerequisite & 3.39 & 6.98 & 5.95 & 4.35 \\
Configuration/environment prerequisite & 5.08 & 3.49 & 3.57 & 2.17 \\
Deployment/external dependency & 3.39 & 5.81 & 2.38 & 1.09 \\

\midrule
Report ambiguity / insufficient detail & 5.08 & 5.81 & 5.95 & 3.26 \\

\midrule
Budget exceeded (step/timeout) & 2.54 & 8.14 & 7.14 & 6.52 \\

\midrule
\textbf{\#Failure labels} & \textbf{118} & \textbf{86} & \textbf{84} & \textbf{92} \\
\bottomrule
\end{tabularx}
\caption{RQ3 failure taxonomy across projects. Percentages are computed over all failure labels in each project.}
\label{tab:rq3-failure-taxonomy}
\end{table}

Table~\ref{tab:rq3-failure-taxonomy} summarizes the distribution of failure labels across projects. \emph{Element grounding or interaction blocked} is the dominant failure mode in every subject, accounting for 60.47\%-73.91\% of all failure labels. These failures occur when the agent cannot bind an intended action to a visible and interactable UI target,  due to overlays, hidden or stale elements, blocked interactions, or dynamic UI updates. This result suggests that, even with live browser-state grounding, robustly mapping report-derived intents to dynamic web UI elements remains a central challenge.

% Failures related to \emph{file/artifact prerequisites, data/database prerequisites, configuration/environment prerequisites, and deployment/external dependencies} also appear consistently across projects, reflecting implicit assumptions in real-world bug reports about execution context. \emph{Budget-exceeded failures} are relatively more frequent on Metabase and NocoDB, suggesting that longer interaction chains or slower state transitions can exceed predefined step or time budgets. Lastly, \emph{report ambiguity or insufficient detail} may prevent the system from establishing a predictable execution path, despite a reasonable amount of exploration. This highlights the influence of report quality on automated reproduction.

Other failures mainly reflect missing execution context or long interaction chains. File/artifact, data/database, configuration/environment, and deployment/external-dependency failures show that real reports often assume resources or states that are not present in the controlled deployment. Budget-exceeded failures are more frequent on Metabase and NocoDB, suggesting that longer workflows or slower state transitions can exceed the step or timeout limits. Report ambiguity also prevents a reliable execution path when the report lacks sufficient detail.

Overall, RQ3 demonstrates that reproduction failures are not primarily caused by system crashes or incorrect action selection, but rather by constraints in the reconstruction of execution prerequisites and the grounding of actions to dynamic web interfaces. These findings indicate that future reproduction systems would be enhanced by more robust UI interaction modeling, environment-aware execution, and a more seamless integration between report comprehension and execution planning.

\begin{tcolorbox}[boxsep=2pt,left=2pt,right=2pt,top=1pt,bottom=1pt, before skip=10pt, after skip=5pt]
\noindent\textbf{Answer to RQ3.} 
% Most failures stem from \textbf{element grounding or blocked interactions}, accounting for roughly \textbf{60\%--74\%} of failure labels across subjects. The remaining failures mainly relate to budgets and missing prerequisites, indicating that robust grounding under dynamic states and prerequisite recovery are the key improvement directions.
Reproduction failures are driven primarily by difficulties in grounding actions to dynamic web UI interfaces (60\%-74\%), rather than incorrect high-level action selection, with secondary failures emphasizing the importance of prerequisite recovery and environment-aware execution.
\end{tcolorbox}

%% file: 5.1-discussion.tex
\section{Discussion}\label{sec:discussion}

\noindent\textbf{Relation to prior bug reproduction systems.}
\name is related to Android bug reproduction systems such as RepRev, ReCDroid+, and ReBL~\cite{li2020automated,zhao2022recdroid+,wang2024feedback}, which reproduce mobile bugs by extracting useful information from reviews or bug reports, matching it to Android GUI elements, and guiding exploration or LLM-based replay. \name shares the same goal of reducing manual reproduction effort, but targets a different setting: web GUI bugs executed through a general-purpose browser. Unlike Android systems that can rely on APKs, Android UI hierarchies, platform instrumentation, and crash signals, \name must reason over browser-observable evidence such as dynamic DOM states, URLs, screenshots, asynchronous page updates, and web-specific artifacts.

\vspace{3pt}
\noindent \name also differs from recent general bug reproduction agents such as AEGIS~\cite{wang2024aegis}, which focus on generating code-level reproduction scripts from issue descriptions. In contrast, \name executes deployed web applications through browser interactions and checks the final browser-observable state against report-derived expectations. Rather than replacing Android or code-level reproduction systems, \name provides a complementary exploration of web GUI bug reproduction, where missing execution context, dynamic user-facing workflows, and browser-observable outcomes play an important role.

%% file: 6-Threats.tex
\section{Threats to Validity}\label{sec:threats}

\noindent\textit{Construct Validity.} 
A primary threat arises from how the reproduction success rate is defined and measured. \name evaluates browser states, such as DOM snapshots, screenshots, and execution traces, against expectations based on bug reports to determine success. This procedure may not capture all subtle or context-dependent signs of a bug, particularly those involving complex visual effects or timing-sensitive interactions. Furthermore, the system's capacity to reproduce the reported failure may be impacted by the quality and completeness of project metadata and supplementary artifacts, which are essential for context reconstruction and state-aware execution.

% \vspace{0.5em}
\noindent\textit{Internal Validity.}
\name depends on large language models to interpret natural-language bug reports and generate executable interaction sequences. Variability in reproduced behavior across runs or model configurations is introduced by the probabilistic nature of LLM outputs. The reproducibility of results may be influenced by prompt design or integration with the browser automation environment, which could also affect performance. To mitigate these effects, evaluations are conducted using consistent model configurations, and each bug report involves multiple reproduction attempts.

% \vspace{0.5em}
\noindent\textit{External Validity.}
Our evaluation is conducted on a curated set of real-world bug reports from selected web applications. While the dataset includes diverse UI patterns and interaction sequences, the findings may not generalize to all web applications, particularly highly dynamic, domain-specific, or rapidly evolving systems. The dataset may also be biased toward more easily reproducible cases due to filtering and environment compatibility, which could affect the generalization of reproduction results to arbitrary bug reports. 
% The main evaluation is conducted on controlled current deployments rather than reconstructing historical buggy snapshots for all 667 reports. This design supports scale and consistent comparison, but some original bug symptoms may not be observable in the evaluated deployments. We mitigate this threat by adding a project-balanced historical replay subset. Since this subset is sampled from successful current-deployment runs, it should be interpreted as complementary validation of successful procedures rather than an unbiased estimate of historical replay success over all reports.main evaluation current deployments.

%% file: 7-Related_work.tex
\section{Related Work}\label{sec:related}

\noindent\textit{Web GUI Testing and Bug Reproduction.}
Prior research on web GUI testing has produced a wide range of automation frameworks and tools, including record and replay systems, script based testing frameworks, and model based approaches~\cite{borjesson2012automated, memon2013automated, banerjee2013graphical, burg2013interactive, artzi2010finding, artzi2011framework, lebeau2013model, mattiello2022model}. Tools such as Selenium and Playwright enable automated browser interaction through manually authored scripts, while model based testing techniques infer UI models to generate test sequences automatically~\cite{holmes2006automating, thooriqoh2021selenium, gojare2015analysis, garcia2024exploring}. Although these methods are effective in regression testing, they are typically limited in their ability to directly reproduce failures from natural language bug reports due to their reliance on predefined test cases, test scripts, or manually specified preconditions~\cite{chaparro2019assessing}.

Several studies have investigated automated bug reproduction, primarily in the context of desktop or mobile applications, where execution traces, stack traces, or system logs are leveraged to guide reproduction~\cite{li2020automated, wang2024feedback, herbold2011improved, zhang2023automatically, johnson2022empirical}. In the web domain, reproduction support remains more limited, as failures often depend on complex UI interactions, dynamic content, and evolving client side state~\cite{di2003considering, mcallister2008leveraging, apaolaza2017wevquery, artzi2010finding}. Consequently, existing approaches typically do not provide end-to-end support for reconstructing reproduction scenarios directly from informal bug descriptions.

% \vspace{0.5em}
\noindent\textit{Large Language Models for Web GUI Testing.}
Recent advances in large language models have stimulated growing interest in applying LLMs to software engineering tasks, including test generation, program repair, code summarization, and bug report analysis~\cite{chen2024chatunitest, schafer2023empirical, yu2023llm, jin2023inferfix, kang2023large, acharya2025can}. Several studies explore the use of LLMs to generate UI test scripts from natural language instructions or high level specifications, and others investigate autonomous web agents capable of performing goal driven interactions~\cite{ran2024guardian, liu2024make, duan2023towards}. While promising, these approaches often focus on instruction following or task completion and provide limited support for reproduction specific requirements, such as context reconstruction, state tracking, and outcome validation grounded in reported failure descriptions~\cite{nashid2025issue2test}. LLM-based web agents further demonstrate autonomous browser interaction~\cite{yang2024agentoccam, deng2023mind2web, zhou2023webarena, ma2023laser, koh2024visualwebarena}, but they are generally optimized for directed task completion rather than report-specific reproduction with prerequisite recovery and oracle-based validation.
% AEGIS also starts from issue descriptions, but it generates code-context reproduction scripts rather than driving live Web GUIs, and its publicly available artifacts do not directly support our browser-based benchmark~\cite{wang2024aegis}.

% \vspace{0.5em}
% \noindent\textit{LLM based Agents.}
% Current research has investigated the use of LLM based agents for software engineering tasks, such as test generation and autonomous web interaction~\cite{yang2024agentoccam, deng2023mind2web, zhou2023webarena, ma2023laser, koh2024visualwebarena}.  Although these agents exhibit the potential of LLMs to connect informal specifications and executable workflows, they are typically designed for directed task completion rather than bug reproduction. 
% They therefore provide limited solutions for reproduction specific problems, like reasoning over dynamic user interface states, determining missing execution context, or verifying results against expectations derived from reports. 

Our work, in contrast, explicitly integrates context reconstruction, state aware execution, and outcome validation to perform end-to-end reproduction of web GUI bugs directly from natural language bug reports, and treats bug reports as actionable inputs.

%% file: 8-Conclusion.tex
\section{Conclusion}\label{sec:conclusion}

We presented \name, an LLM-driven browser agent for reconstructing and validating browser reproduction procedures from web GUI bug reports. \name combines prerequisite preparation, state-aware browser execution, and oracle-based final-state validation. On 667 bug reports from four open source web applications, \name achieves an average RSR of 49.96\% and outperforms both direct script generation and a browser-agent baseline. A historical replay study on 40 successful cases further shows that the reconstructed procedures can usually trigger the original bug-present behavior when the corresponding buggy versions are restored. The ablation study shows that context reconstruction improves success, while failure analysis identifies UI grounding on dynamic web interfaces as the dominant bottleneck. Future work will improve grounding for complex widgets, prerequisite recovery, and the integration of reproduction traces with debugging and regression testing.

%% file: 9-References.bib
@article{brisset2022erratum,
  title={Erratum: Leveraging flexible tree matching to repair broken locators in web automation scripts},
  author={Brisset, Sacha and Rouvoy, Romain and Seinturier, Lionel and Pawlak, Renaud},
  journal={Information and Software Technology},
  volume={144},
  pages={106754},
  year={2022},
  publisher={Elsevier}
}

@article{deng2023mind2web,
  title={Mind2web: Towards a generalist agent for the web},
  author={Deng, Xiang and Gu, Yu and Zheng, Boyuan and Chen, Shijie and Stevens, Sam and Wang, Boshi and Sun, Huan and Su, Yu},
  journal={Advances in Neural Information Processing Systems},
  volume={36},
  pages={28091--28114},
  year={2023}
}

@article{zhou2023webarena,
  title={Webarena: A realistic web environment for building autonomous agents},
  author={Zhou, Shuyan and Xu, Frank F and Zhu, Hao and Zhou, Xuhui and Lo, Robert and Sridhar, Abishek and Cheng, Xianyi and Ou, Tianyue and Bisk, Yonatan and Fried, Daniel and others},
  journal={arXiv preprint arXiv:2307.13854},
  year={2023}
}

@article{zheng2024gpt,
  title={Gpt-4v (ision) is a generalist web agent, if grounded},
  author={Zheng, Boyuan and Gou, Boyu and Kil, Jihyung and Sun, Huan and Su, Yu},
  journal={arXiv preprint arXiv:2401.01614},
  year={2024}
}

@inproceedings{yao2022react,
  title={React: Synergizing reasoning and acting in language models},
  author={Yao, Shunyu and Zhao, Jeffrey and Yu, Dian and Du, Nan and Shafran, Izhak and Narasimhan, Karthik R and Cao, Yuan},
  booktitle={The eleventh international conference on learning representations},
  year={2022}
}

@article{lota5060080recent,
  title={Recent Trends and Challenges in Using Nlp Techniques in Software Debugging: A Systematic Literature Review},
  author={Lota, Lutfun Nahar and Zaman, Tarannum Shaila and Azwad, Mirza Mohammad and Farah, Labiba and Chowdhury, Abrar and Anjum, Zaarin and Islam, Chadni and Kamal, Abu Raihan Mostofa},
  journal={Available at SSRN 5060080}
}

@article{wang2024application,
  title={Application Monitoring for bug reproduction in web-based applications},
  author={Wang, Di and Galster, Matthias and Morales-Trujillo, Miguel},
  journal={Journal of Systems and Software},
  volume={207},
  pages={111834},
  year={2024},
  publisher={Elsevier}
}

@inproceedings{li2020automated,
  title={Automated bug reproduction from user reviews for android applications},
  author={Li, Shuyue and Guo, Jiaqi and Fan, Ming and Lou, Jian-Guang and Zheng, Qinghua and Liu, Ting},
  booktitle={Proceedings of the ACM/IEEE 42nd International Conference on Software Engineering: Software Engineering in Practice},
  pages={51--60},
  year={2020}
}

@article{zhao2022recdroid+,
  title={Recdroid+: Automated end-to-end crash reproduction from bug reports for android apps},
  author={Zhao, Yu and Su, Ting and Liu, Yang and Zheng, Wei and Wu, Xiaoxue and Kavuluru, Ramakanth and Halfond, William GJ and Yu, Tingting},
  journal={ACM Transactions on Software Engineering and Methodology (TOSEM)},
  volume={31},
  number={3},
  pages={1--33},
  year={2022},
  publisher={ACM New York, NY}
}

@inproceedings{yu2023llm,
  title={Llm for test script generation and migration: Challenges, capabilities, and opportunities},
  author={Yu, Shengcheng and Fang, Chunrong and Ling, Yuchen and Wu, Chentian and Chen, Zhenyu},
  booktitle={2023 IEEE 23rd International Conference on Software Quality, Reliability, and Security (QRS)},
  pages={206--217},
  year={2023},
  organization={IEEE}
}

@inproceedings{kang2023large,
  title={Large language models are few-shot testers: Exploring llm-based general bug reproduction},
  author={Kang, Sungmin and Yoon, Juyeon and Yoo, Shin},
  booktitle={2023 IEEE/ACM 45th International Conference on Software Engineering (ICSE)},
  pages={2312--2323},
  year={2023},
  organization={IEEE}
}

@article{khatib2025assertflip,
  title={AssertFlip: Reproducing Bugs via Inversion of LLM-Generated Passing Tests},
  author={Khatib, Lara and Mathews, Noble Saji and Nagappan, Meiyappan},
  journal={arXiv preprint arXiv:2507.17542},
  year={2025}
}

@article{chen2025standing,
  title={Standing on the Shoulders of Giants: Bug-Aware Automated GUI Testing via Retrieval Augmentation},
  author={Chen, Mengzhuo and Liu, Zhe and Chen, Chunyang and Wang, Junjie and Wu, Boyu and Hu, Jun and Wang, Qing},
  journal={Proceedings of the ACM on Software Engineering},
  volume={2},
  number={FSE},
  pages={825--846},
  year={2025},
  publisher={ACM New York, NY, USA}
}

@inproceedings{roehm2013monitoring,
  title={Monitoring user interactions for supporting failure reproduction},
  author={Roehm, Tobias and Gurbanova, Nigar and Bruegge, Bernd and Joubert, Christophe and Maalej, Walid},
  booktitle={2013 21st International Conference on Program Comprehension (ICPC)},
  pages={73--82},
  year={2013},
  organization={IEEE}
}

@inproceedings{burg2013interactive,
  title={Interactive record/replay for web application debugging},
  author={Burg, Brian and Bailey, Richard and Ko, Amy J and Ernst, Michael D},
  booktitle={Proceedings of the 26th annual ACM symposium on User interface software and technology},
  pages={473--484},
  year={2013}
}

@inproceedings{herbold2011improved,
  title={Improved bug reporting and reproduction through non-intrusive gui usage monitoring and automated replaying},
  author={Herbold, Steffen and Grabowski, Jens and Waack, Stephan and B{\"u}nting, Uwe},
  booktitle={2011 IEEE fourth international conference on software testing, verification and validation workshops},
  pages={232--241},
  year={2011},
  organization={IEEE}
}

@inproceedings{bettenburg2008makes,
  title={What makes a good bug report?},
  author={Bettenburg, Nicolas and Just, Sascha and Schr{\"o}ter, Adrian and Weiss, Cathrin and Premraj, Rahul and Zimmermann, Thomas},
  booktitle={Proceedings of the 16th ACM SIGSOFT International Symposium on Foundations of software engineering},
  pages={308--318},
  year={2008}
}

@inproceedings{zhou2012should,
  title={Where should the bugs be fixed? more accurate information retrieval-based bug localization based on bug reports},
  author={Zhou, Jian and Zhang, Hongyu and Lo, David},
  booktitle={2012 34th International conference on software engineering (ICSE)},
  pages={14--24},
  year={2012},
  organization={IEEE}
}

@inproceedings{chaparro2019assessing,
  title={Assessing the quality of the steps to reproduce in bug reports},
  author={Chaparro, Oscar and Bernal-C{\'a}rdenas, Carlos and Lu, Jing and Moran, Kevin and Marcus, Andrian and Di Penta, Massimiliano and Poshyvanyk, Denys and Ng, Vincent},
  booktitle={Proceedings of the 2019 27th ACM joint meeting on european software engineering conference and symposium on the foundations of software engineering},
  pages={86--96},
  year={2019}
}

@article{gojare2015analysis,
  title={Analysis and design of selenium webdriver automation testing framework},
  author={Gojare, Satish and Joshi, Rahul and Gaigaware, Dhanashree},
  journal={Procedia Computer Science},
  volume={50},
  pages={341--346},
  year={2015},
  publisher={Elsevier}
}

@book{pathak2024web,
  title={Web Automation Testing Using Playwright: End-to-end, API, accessibility, and visual testing using Playwright},
  author={Pathak, Kailash},
  year={2024},
  publisher={BPB Publications}
}

@inproceedings{memon2013automated,
  title={Automated testing of GUI applications: models, tools, and controlling flakiness},
  author={Memon, Atif M and Cohen, Myra B},
  booktitle={2013 35th International Conference on Software Engineering (ICSE)},
  pages={1479--1480},
  year={2013},
  organization={IEEE}
}

@article{banerjee2013graphical,
  title={Graphical user interface (GUI) testing: Systematic mapping and repository},
  author={Banerjee, Ishan and Nguyen, Bao and Garousi, Vahid and Memon, Atif},
  journal={Information and Software Technology},
  volume={55},
  number={10},
  pages={1679--1694},
  year={2013},
  publisher={Elsevier}
}

@inproceedings{borjesson2012automated,
  title={Automated system testing using visual gui testing tools: A comparative study in industry},
  author={Borjesson, Emil and Feldt, Robert},
  booktitle={2012 IEEE Fifth International Conference on Software Testing, Verification and Validation},
  pages={350--359},
  year={2012},
  organization={IEEE}
}

@article{artzi2010finding,
  title={Finding bugs in web applications using dynamic test generation and explicit-state model checking},
  author={Artzi, Shay and Kiezun, Adam and Dolby, Julian and Tip, Frank and Dig, Daniel and Paradkar, Amit and Ernst, Michael D},
  journal={IEEE Transactions on Software Engineering},
  volume={36},
  number={4},
  pages={474--494},
  year={2010},
  publisher={IEEE}
}

@inproceedings{artzi2011framework,
  title={A framework for automated testing of JavaScript web applications},
  author={Artzi, Shay and Dolby, Julian and Jensen, Simon Holm and M{\o}ller, Anders and Tip, Frank},
  booktitle={Proceedings of the 33rd international conference on software engineering},
  pages={571--580},
  year={2011}
}

@inproceedings{lebeau2013model,
  title={Model-based vulnerability testing for web applications},
  author={Lebeau, Franck and Legeard, Bruno and Peureux, Fabien and Vernotte, Alexandre},
  booktitle={2013 IEEE Sixth International Conference on Software Testing, Verification and Validation Workshops},
  pages={445--452},
  year={2013},
  organization={IEEE}
}

@article{mattiello2022model,
  title={Model-based testing leveraged for automated web tests},
  author={Mattiello, Guilherme Ricken and Endo, Andr{\'e} Takeshi},
  journal={Software Quality Journal},
  volume={30},
  number={3},
  pages={621--649},
  year={2022},
  publisher={Springer}
}

@article{thooriqoh2021selenium,
  title={Selenium framework for web automation testing: A systematic literature review},
  author={Thooriqoh, Hazna At and Annisa, Tiara Nur and Yuhana, Umi Laili},
  journal={JUTI: Jurnal Ilmiah Teknologi Informasi},
  pages={65--76},
  year={2021}
}

@inproceedings{holmes2006automating,
  title={Automating functional tests using selenium},
  author={Holmes, Antawan and Kellogg, Marc},
  booktitle={AGILE 2006 (AGILE'06)},
  pages={6--pp},
  year={2006},
  organization={IEEE}
}

@inproceedings{garcia2024exploring,
  title={Exploring browser automation: A comparative study of selenium, cypress, puppeteer, and playwright},
  author={Garc{\'\i}a, Boni and del Alamo, Jose M and Leotta, Maurizio and Ricca, Filippo},
  booktitle={International Conference on the Quality of Information and Communications Technology},
  pages={142--149},
  year={2024},
  organization={Springer}
}

@inproceedings{wang2024feedback,
  title={Feedback-driven automated whole bug report reproduction for android apps},
  author={Wang, Dingbang and Zhao, Yu and Feng, Sidong and Zhang, Zhaoxu and Halfond, William GJ and Chen, Chunyang and Sun, Xiaoxia and Shi, Jiangfan and Yu, Tingting},
  booktitle={Proceedings of the 33rd ACM SIGSOFT International Symposium on Software Testing and Analysis},
  pages={1048--1060},
  year={2024}
}

@inproceedings{zhang2023automatically,
  title={Automatically reproducing android bug reports using natural language processing and reinforcement learning},
  author={Zhang, Zhaoxu and Winn, Robert and Zhao, Yu and Yu, Tingting and Halfond, William GJ},
  booktitle={Proceedings of the 32nd ACM SIGSOFT International Symposium on Software Testing and Analysis},
  pages={411--422},
  year={2023}
}

@inproceedings{mcallister2008leveraging,
  title={Leveraging user interactions for in-depth testing of web applications},
  author={McAllister, Sean and Kirda, Engin and Kruegel, Christopher},
  booktitle={International Workshop on Recent Advances in Intrusion Detection},
  pages={191--210},
  year={2008},
  organization={Springer}
}

@article{apaolaza2017wevquery,
  title={WevQuery: Testing hypotheses about web interaction patterns},
  author={Apaolaza, Aitor and Vigo, Markel},
  journal={Proceedings of the ACM on Human-Computer Interaction},
  volume={1},
  number={EICS},
  pages={1--17},
  year={2017},
  publisher={ACM New York, NY, USA}
}

@inproceedings{di2003considering,
  title={Considering browser interaction in web application testing},
  author={Di Lucca, Giuseppe A and Di Penta, Massimiliano},
  booktitle={Fifth IEEE International Workshop on Web Site Evolution, 2003. Theme: Architecture. Proceedings.},
  pages={74--81},
  year={2003},
  organization={IEEE}
}

@inproceedings{chen2024chatunitest,
  title={Chatunitest: A framework for llm-based test generation},
  author={Chen, Yinghao and Hu, Zehao and Zhi, Chen and Han, Junxiao and Deng, Shuiguang and Yin, Jianwei},
  booktitle={Companion Proceedings of the 32nd ACM International Conference on the Foundations of Software Engineering},
  pages={572--576},
  year={2024}
}

@article{schafer2023empirical,
  title={An empirical evaluation of using large language models for automated unit test generation},
  author={Sch{\"a}fer, Max and Nadi, Sarah and Eghbali, Aryaz and Tip, Frank},
  journal={IEEE Transactions on Software Engineering},
  volume={50},
  number={1},
  pages={85--105},
  year={2023},
  publisher={IEEE}
}

@inproceedings{jin2023inferfix,
  title={Inferfix: End-to-end program repair with llms},
  author={Jin, Matthew and Shahriar, Syed and Tufano, Michele and Shi, Xin and Lu, Shuai and Sundaresan, Neel and Svyatkovskiy, Alexey},
  booktitle={Proceedings of the 31st ACM joint european software engineering conference and symposium on the foundations of software engineering},
  pages={1646--1656},
  year={2023}
}

@inproceedings{acharya2025can,
  title={Can we enhance bug report quality using llms?: An empirical study of llm-based bug report generation},
  author={Acharya, Jagrit and Ginde, Gouri},
  booktitle={Proceedings of the 29th International Conference on Evaluation and Assessment in Software Engineering},
  pages={994--1003},
  year={2025}
}

@inproceedings{ran2024guardian,
  title={Guardian: A runtime framework for LLM-based UI exploration},
  author={Ran, Dezhi and Wang, Hao and Song, Zihe and Wu, Mengzhou and Cao, Yuan and Zhang, Ying and Yang, Wei and Xie, Tao},
  booktitle={Proceedings of the 33rd ACM SIGSOFT International Symposium on Software Testing and Analysis},
  pages={958--970},
  year={2024}
}

@inproceedings{liu2024make,
  title={Make llm a testing expert: Bringing human-like interaction to mobile gui testing via functionality-aware decisions},
  author={Liu, Zhe and Chen, Chunyang and Wang, Junjie and Chen, Mengzhuo and Wu, Boyu and Che, Xing and Wang, Dandan and Wang, Qing},
  booktitle={Proceedings of the IEEE/ACM 46th International Conference on Software Engineering},
  pages={1--13},
  year={2024}
}

@inproceedings{duan2023towards,
  title={Towards generating UI design feedback with LLMs},
  author={Duan, Peitong and Warner, Jeremy and Hartmann, Bjoern},
  booktitle={Adjunct Proceedings of the 36th Annual ACM Symposium on User Interface Software and Technology},
  pages={1--3},
  year={2023}
}

@article{nashid2025issue2test,
  title={Issue2Test: Generating Reproducing Test Cases from Issue Reports},
  author={Nashid, Noor and Bouzenia, Islem and Pradel, Michael and Mesbah, Ali},
  journal={arXiv preprint arXiv:2503.16320},
  year={2025}
}

@inproceedings{johnson2022empirical,
  title={An empirical investigation into the reproduction of bug reports for android apps},
  author={Johnson, Jack and Mahmud, Junayed and Wendland, Tyler and Moran, Kevin and Rubin, Julia and Fazzini, Mattia},
  booktitle={2022 IEEE International Conference on Software Analysis, Evolution and Reengineering (SANER)},
  pages={321--322},
  year={2022},
  organization={IEEE}
}

@article{wang2024aegis,
  title={AEGIS: An agent-based framework for general bug reproduction from issue descriptions},
  author={Wang, Xinchen and Gao, Pengfei and Meng, Xiangxin and Peng, Chao and Hu, Ruida and Lin, Yun and Gao, Cuiyun},
  journal={arXiv preprint arXiv:2411.18015},
  year={2024}
}

@article{yang2024agentoccam,
  title={Agentoccam: A simple yet strong baseline for llm-based web agents},
  author={Yang, Ke and Liu, Yao and Chaudhary, Sapana and Fakoor, Rasool and Chaudhari, Pratik and Karypis, George and Rangwala, Huzefa},
  journal={arXiv preprint arXiv:2410.13825},
  year={2024}
}

@article{ma2023laser,
  title={Laser: Llm agent with state-space exploration for web navigation},
  author={Ma, Kaixin and Zhang, Hongming and Wang, Hongwei and Pan, Xiaoman and Yu, Wenhao and Yu, Dong},
  journal={arXiv preprint arXiv:2309.08172},
  year={2023}
}

@inproceedings{koh2024visualwebarena,
  title={Visualwebarena: Evaluating multimodal agents on realistic visual web tasks},
  author={Koh, Jing Yu and Lo, Robert and Jang, Lawrence and Duvvur, Vikram and Lim, Ming and Huang, Po-Yu and Neubig, Graham and Zhou, Shuyan and Salakhutdinov, Russ and Fried, Daniel},
  booktitle={Proceedings of the 62nd Annual Meeting of the Association for Computational Linguistics (Volume 1: Long Papers)},
  pages={881--905},
  year={2024}
}

@inproceedings{feng2024prompting,
  title={Prompting is all you need: Automated android bug replay with large language models},
  author={Feng, Sidong and Chen, Chunyang},
  booktitle={Proceedings of the 46th IEEE/ACM International Conference on Software Engineering},
  pages={1--13},
  year={2024}
}

@inproceedings{chaparro2017detecting,
  title={Detecting missing information in bug descriptions},
  author={Chaparro, Oscar and Lu, Jing and Zampetti, Fiorella and Moreno, Laura and Di Penta, Massimiliano and Marcus, Andrian and Bavota, Gabriele and Ng, Vincent},
  booktitle={Proceedings of the 2017 11th joint meeting on foundations of software engineering},
  pages={396--407},
  year={2017}
}

@inproceedings{davies2014s,
  title={What's in a bug report?},
  author={Davies, Steven and Roper, Marc},
  booktitle={Proceedings of the 8th ACM/IEEE International Symposium on Empirical Software Engineering and Measurement},
  pages={1--10},
  year={2014}
}

@inproceedings{AnvikHM06,
  author       = {John Anvik and
                  Lyndon Hiew and
                  Gail C. Murphy},
  editor       = {Leon J. Osterweil and
                  H. Dieter Rombach and
                  Mary Lou Soffa},
  title        = {Who should fix this bug?},
  booktitle    = {28th International Conference on Software Engineering {(ICSE} 2006),
                  Shanghai, China, May 20-28, 2006},
  pages        = {361--370},
  publisher    = {{ACM}},
  year         = {2006},
}
